\documentclass[12pt]{emulateapj}
\usepackage{graphicx}
\usepackage{times}
\usepackage{natbib}
\usepackage{amsfonts}
\usepackage{amsmath}
\usepackage{amsbsy}
\usepackage{bm}
\usepackage{hyperref}
\usepackage{url}
\usepackage{microtype}
\usepackage{rotating}
\usepackage{booktabs}
\usepackage{threeparttable}
\usepackage{tabularx}
\usepackage{subfigure}

\newcommand{\project}[1]{\textsl{#1}}
\newcommand{\fermi}{\project{Fermi}}
\newcommand{\rxte}{\project{RXTE}}
\newcommand{\given}{\,|\,}
\newcommand{\dd}{\mathrm{d}}
\newcommand{\counts}{y}

\newcommand{\mean}{\lambda}
\newcommand{\likelihood}{{\mathcal L}}

\newcommand{\bg}{\mathrm{bg}}
\newcommand{\word}{\phi}

\begin{document}

\title{Dissecting Magnetar Variability with Bayesian Hierarchical Models}

\author{Daniela Huppenkothen\altaffilmark{1, 2, 3, 4}, Brendon J. Brewer\altaffilmark{5}, David W. Hogg\altaffilmark{3,2}, Iain Murray\altaffilmark{6}, Marcus Frean\altaffilmark{7}, Chris Elenbaas\altaffilmark{1}, Anna L. Watts\altaffilmark{1}, Yuri Levin\altaffilmark{8}, Alexander J. van der Horst\altaffilmark{1}, Chryssa Kouveliotou\altaffilmark{9,10}}
 
\altaffiltext{1}{Anton Pannekoek Institute for Astronomy, University of
  Amsterdam, Postbus 94249, 1090 GE Amsterdam, the Netherlands}
  \altaffiltext{2}{Center for Data Science, New York University, 726 Broadway, 7th Floor, New York, NY 10003}
  \altaffiltext{3}{Center for Cosmology and Particle Physics, Department of Physics, New York University, 4 Washington Place, New York, NY 10003, USA}
  \altaffiltext{4}{E-mail: daniela.huppenkothen@nyu.edu}
  \altaffiltext{5}{Department of Statistics, The University of Auckland, Private Bag 92019, Auckland 1142, New Zealand}
\altaffiltext{6}{School of Informatics, University of Edinburgh}
\altaffiltext{7}{School of Engineering and Computer Science, Victoria University of Wellington, New Zealand}
\altaffiltext{8}{Monash Center for Astrophysics and School of Physics, Monash University, Clayton, Victoria 3800, Australia}
\altaffiltext{9}{Astrophysics Office, ZP 12, NASA/Marshall Space Flight Center, Huntsville, AL 35812, USA}
\altaffiltext{10}{NSSTC, 320 Sparkman Drive, Huntsville, AL 35805, USA}

\begin{abstract}
Neutron stars are a prime laboratory for testing physical processes under conditions of strong gravity, high density, and extreme magnetic fields.
Among the zoo of neutron star phenomena, magnetars stand out for their bursting behaviour, ranging from extremely bright, rare giant flares to numerous, 
less energetic recurrent bursts. The exact trigger and emission mechanisms for these bursts are not known; favoured models involve either a crust fracture
and subsequent energy release into the magnetosphere, or explosive reconnection of magnetic field lines. 
In the absence of a predictive model, understanding the physical processes responsible for
magnetar burst variability is difficult. Here, we 
develop an empirical model that decomposes magnetar bursts into a superposition of small spike-like features with a simple functional form, where the number 
of model components is itself part of the inference problem.
The cascades of spikes that we model might be formed by avalanches of reconnection, or crust rupture aftershocks.
Using Markov Chain Monte Carlo (MCMC) sampling augmented with reversible jumps between models with different numbers of parameters, we characterise the posterior distributions of the model 
parameters and the number of components per burst. We relate these model parameters to physical quantities in the system, and show for the first time that 
the variability within a burst does not conform to predictions from ideas of self-organised criticality.  We also examine how well the properties of the spikes fit the predictions of simplified cascade models for the different trigger mechanisms.  

\end{abstract}

\keywords{pulsars: individual (SGR J1550-5418), stars: magnetic fields, stars: neutron, X-rays: bursts, methods:statistics}

\section{Introduction}

With current and upcoming telescopes monitoring the sky regularly across the entire electromagnetic spectrum, time domain astronomy is emerging as one of the key fields in which major 
new discoveries are being made.  A large fraction of astrophysical sources are known to be variable. The timescales span more than five orders of magnitude: fast oscillations in 
X-ray binaries (XRBs) change over milliseconds \citep[e.g.][]{xrb_khzqpos}, while red giants have been observed to change over decades or even centuries \citep[e.g.][]{dasch_giants}. 
Variability studies have the potential to unravel fundamental physical processes: studying variability in XRBs can help us unravel accretion processes and constrain theories of viscosity and strong gravity. 
Similarly, giant flares from magnetars have the potential to map the neutron star interior via neutron star seismology \citep[e.g.\ ][]{steiner2009}. 

\begin{figure*}[htbp]
\begin{center}
\includegraphics[width=\textwidth]{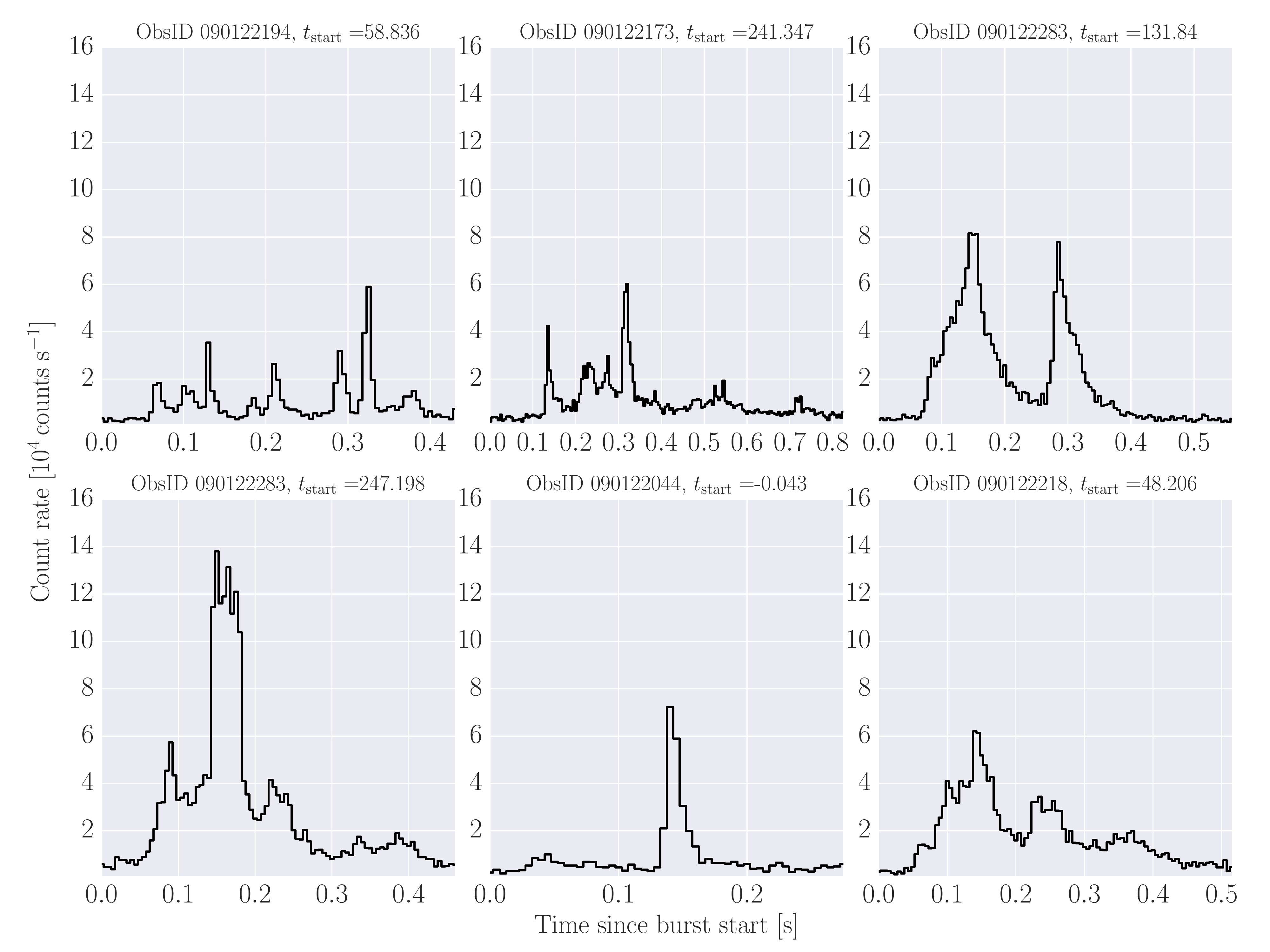}
\caption{Six example bursts observed with \fermi/GBM from the magnetar SGR J1550-5418. The light curves have a time resolution of $0.005\,\mathrm{s}$. Bursts show a complex morphology of one or more
spikes that differs drastically from burst to burst, making development of a common model for all burst light curves difficult. For details of the source
and this sample, see Section \ref{ch6:data}.}
\label{fig:example_bursts}
\end{center}
\end{figure*}

While much of the variability exhibited for example in XRBs can be characterised using standard Fourier methodology, these methods are not appropriate for an important group of sources: transients 
with complex temporal structure that is a superposition of different processes. When attempting to disentangle the individual components (for example, a periodic signal from underlying stochastic variability)
using standard methods, it is possible to introduce systematic errors into inferences made from this type of data. Three examples stand out particularly: solar flares, $\gamma$-ray bursts (GRBs) and magnetar bursts.  All three phenomena are characterised by bursts lasting from $\sim\!\! 1/10$ of a second to hundreds of seconds, and a 
complex temporal structure that varies strongly from burst to burst (for an example, see Figure \ref{fig:example_bursts}). 

While Fourier transforming the data is always possible, it is not necessarily the best approach to characterising variability: inferences are most reliable when trying to find
periodic signals in a background that is straightforward to characterise (in particular, a stationary stochastic background).
Fourier methods fail in particular when the underlying processes contributing to the observed light curve, especially deterministic 
components in combination with stochastic variability, are not well known. In this case,
the statistical distributions in the periodogram are not straightforward to model, thus making inferences about the variability in the source problematic. Conversely, 
it is difficult to postulate a common model applicable to a large sample of light curves of these sources. Any light curve model must be flexible enough to account for differences between bursts. 
Here, we aim to develop a flexible, generative 
model for highly variable transient events, based on a decomposition of the light curve into simple shapes, without knowing the number of components in the model a priori. We demonstrate the 
power of this approach on a large sample of magnetar bursts, and, for the first time, connect variability in magnetar bursts to the time scales thought to govern the underlying physics.


Neutron stars, the ultra-dense compact remnants of core-collapse supernovae, are the prime laboratory for studying nuclear 
physical processes in a parameter regime of density and pressure inaccessible to experiments on Earth. 
Among the veritable zoo of neutron stars, two historically separate groups stood out for their peculiar properties: Soft Gamma Repeaters (SGRs), named after their frequently recurring bursts in hard X-rays, and Anomalous X-ray Pulsars (AXPs), isolated from the bulk of the X-ray pulsars based on their persistent X-ray
emission characteristics. In the last decade, however, both groups have been found to form a single class of neutron stars with extreme magnetic fields, generally called {\it magnetars} \citep{duncan1992,thompson1995,kouveliotou1998}. In the original scenario, the observable X-ray emission, both persistent emission and bursts, 
is powered by the slow evolution and decay of the source's strong magnetic field, instead of the loss of rotational energy as is generally the case for standard radio pulsars \citep{thompson1995,thompson2001}. 

Most magnetars share common properties \citep[for general overviews, see ][]{woods2006,mereghetti2011}: slow spin periods between
$2$--$12\,\mathrm{s}$, which, coupled with generally high period derivatives, lead to large inferred dipole magnetic fields of
the order of $10^{14} \, \mathrm{G}$, well above the quantum-critical limit $B_{\mathrm{QED}} = 4.4 \times 10^{13} \, \mathrm{G}$,
where quantum effects such as pair production and photon splitting become important \citep[although three sources have been 
identified with properties similar to magnetars, but with inferred dipole fields below this limit;][]{vanderhorst2010,esposito2010,rea2010,rea2012,scholz2012,rea2014}. 

Magnetar bursts are of particular interest for a variety of reasons. While the most spectacular outbursts are the rare, but extremely energetic giant flares,
magnetars also show a complex behaviour of emitting much smaller, shorter recurrent bursts. These bursts have a duration of $\lesssim 1\,\mathrm{s}$, with energies
generally between $10^{38}\,\mathrm{erg}$ and $10^{41}\,\mathrm{erg}$, and have a complex temporal structure with single or
multiple peaks that differs from one burst to the next. They are observed either appearing individually, or in burst 
storms, where tens or hundreds of bursts can occur over a timescale of single days to weeks \citep{mazets1999,goetz2006b,israel2008,mereghetti2009,savchenko2010,israel2010,scholz2011,dib2012,vanderhorst2012,vonkienlin2012}. 
The appearance of bursts appears to be random \citep{gogus1999,gogus2000}, but far more numerous than the giant flares: 
for the two best-observed magnetars, SGR 1806-20 and SGR 1900+14, their
data set spans thousands of such bursts. 

One observation of particular interest concerns the overall distributions of these bursts: the differential distribution of fluence (integrated flux) and 
the cumulative distribution of waiting times are similar to those observed in earthquakes and solar flares \citep{cheng1996,gogus1999,gogus2000,prieskorn2012}. The fluence distribution
in particular can be well-modelled by a power law with an index of $\sim\!\! 1.7$, believed to be a typical signature for a system obeying 
self-organised criticality (SOC; \citealp{bak1987,bak1988}; for a recent introduction and review on SOC in astrophysics see \citealp{aschwanden2014}). 
In the SOC framework, the physical system in question continuously drives itself towards a critical state,
without requiring fine tuning. When the critical state is reached, relaxation occurs via a catastrophic release of energy. The system returns
to a subcritical state, and the cycle begins anew. 
One advantage of describing a system in the SOC framework is that while the details of the process leading to the critical state and subsequent
relaxation depend on the physical processes specific to the system, many of the overall statistical properties are universal.

At present, it is not clear whether magnetar bursts are smaller-scale versions of the processes believed to produce a giant flare:
either a rapid rupturing of the neutron star due to an internal re-structuring of the magnetic field, with the energy of the cracking process
released in the magnetosphere, or a slow untwisting of the magnetic field leading to ductile deformations in the crust and explosive 
reconnection in the magnetosphere.

Because of the complex temporal morphology, which differs from burst to burst, it is difficult to extract information from burst light curves. 
Many light curves show a pattern of one or multiple spikes, which are sometimes sitting on top of each other. To extract parameters that could be related
to physical quantities such as the rise time of each spike and the waiting time distribution between spikes, we need to be able to infer both the 
number of spikes per burst as well as the individual parameters of each spike: this requires efficient sampling over a large parameter space.
At the same time, this process needs to operate automatically without the user's intervention, both to avoid biases introduced by manual fitting, as well as
 to allow for an analysis of the large numbers of magnetar bursts observed from various sources.

 Comparable studies abound in the literature on $\gamma$-Ray Bursts (GRBs), where variability has been used to study the details of the physical processes
 involved in the creation of GRB prompt emission in $\gamma$-rays. GRBs have morphologically similar light curves, usually 
 fit with a superposition of pulses, each with an exponential rise and decay as well as a peakedness parameter that parametrizes the roundness of the pulse peak. 
 These studies are generally split into two steps: a pulse-detection step, and an inference step, where detected pulses from many GRBs are combined into an 
 ensemble to infer properties of the entire population of bursts. Many of such studies rely on the visual identification of pulses in the data \citep[e.g.\ ][]{norris1996,norris1999,kocevski2003}
 
 One popular set of pulse detection algorithms involves the identification of time bins with counts that exceed some fixed multiple of the standard deviation of the background and the 
 identification of troughs of emission on either side of such a peak \citep[see e.g.\ ][]{li1996,quilligan2002,guidorzi2015}. 
 Another method first finds rapidly varying time segments in the data using a Bayesian Blocks approach 
\citep{scargle2013}, which are subsequently used as first guessed for fitting a set of pulses to the data, removing the smallest pulses based on some significance criterion \citep{scargle1998,hakkila2011}.
 
 All these approaches have in common that they set a hard limit (usually in peak amplitude, but sometimes in total integrated counts) to what they consider to be a burst, resulting in a set of pulses above this threshold that is then used for the subsequent sample inference. Many of these algorithms are additionally tuned deliberately towards rejecting weak features 
 to keep the number of false positives low. This is equivalent to applying a step-function prior to the relevant pulse 
 parameters (amplitude, or a combination of amplitude and duration). However, because subsequent analyses are based on the ensemble of pulses above the threshold only,
 this prior is never taken into account in the inferences derived from the sample (for a discussion of this problem applied to the waiting time distribution of pulses in GRBs,
 see \citealt{baldeschi2015}). 
 
The purpose of this paper is twofold: in the first part, we propose a similar approach to model magnetar bursts as a linear combination of simple shapes, 
but without enforcing a specific limit on the significance of the detected features. Instead, we allow the model to explore a parameter space  
(in practice, using Markov Chain Monte Carlo [MCMC]) that includes both the pulse parameters for each included model component, as well as 
 the hyper parameters determining the prior distributions of the pulse parameters. We make no attempt to state our confidence in the presence of any 
 given individual model component, but instead work with the posterior distribution rather than a subset of it, and thus include the inherent uncertainty 
 caused by noisy data in our inference over the sample.
 
 Secondly, we present a first, exploratory analysis of the kind of results that may be derived from a model such as the one we present here. In a simple, naive way,
 we characterize differential distributions of key quantities, and find them inconsistent with predictions from self-organized criticality. We search for and find correlations 
 between key physical quantities. This work is only a first step toward a generative model for magnetar bursts, and will be extended to samples of magnetar bursts 
 (and indeed, even over the population of magnetars as a whole) in future work.

In Section \ref{ch6:data}, we present our sample of bursts from SGR J1550-5418, observed with the 
Gamma-ray Burst Monitor (GBM) on board the \fermi\ Gamma-Ray Space Telescope.
The high sensitivity of the instrument as well as 
the large number of observed bursts make this source an excellent target to demonstrate the power of the proposed methods, which we
lay out in detail in Section \ref{ch6:methods}.
In Section \ref{ch6:modelsims}, we test our model on simple simulations of single-spiked bursts to test how well the model recovers the burst parameters, while
in Section \ref{ch6:oneburst}, we show an example of a model for a single burst. Finally, in Section \ref{ch6:results}, we apply our method to a large data set of bursts, 
which allows us to extract a wealth of physically relevant time scales from magnetar bursts. 
We place these timescales in the context of the SOC framework, and tie them to physical parameters in the system in Section \ref{ch6:discussion}.


\section{The Burst Sample}
\label{ch6:data}

SGR J1550-5418 (also 1E 1547.0-5408) was first observed with the {\it Einstein} X-ray observatory \citep{lamb1981}
and subsequently classified as an AXP based on its soft X-ray spectrum and possible association with a supernova remnant \citep{gelfand2007}.
In 2008 and 2009, SGR J1550-5418 exhibited three major bursting episodes (October 2008, January 2009 and March/April 2009), where the 2009 January episode is of special interest 
because it included a very large number of bursts (burst storm). Hundreds of bursts were observed within a single day with various X-ray telescopes: 
the \project{Swift} Burst Alert Telescope (BAT), \citep{israel2010, scholz2011}); \fermi/GBM, \citep{kaneko2010,vonkienlin2012,vanderhorst2012}; the \project{Rossi} 
X-ray Timing Explorer (\rxte), \citep{dib2012}, and the two main instruments on board the \project{INTEGRAL} spacecraft \citep{mereghetti2009, savchenko2010}.
 
Here, we use a sample of bursts observed with \fermi/GBM \citep{meegan2009} during all three bursting episodes. Bursts from SGR J1550-5418 triggered \fermi/GBM for a total of $126$ 
times between 2008 October 3 and 2009 April 17, with $\sim\!\! 450$ bursts observed on its most active day, 2009 January 22, alone. However, not all of these bursts have data with high time resolution
available: each trigger records high time-resolution photon arrival times called time-tagged events, or TTE, at a resolution of $\sim\!\! 2\mu\mathrm{s}$ from $30\,\mathrm{s}$ 
before each trigger to $300\,\mathrm{s}$ after each trigger, upon which the instrument cannot trigger for another $\sim\!\! 300 \,\mathrm{s}$. The data taken in other observing modes in between intervals covered
by individual triggers is not of sufficiently high time resolution to perform detailed timing studies, hence we exclude all bursts without TTE data available.
Within a GBM trigger, subsequent bursts do not trigger the instrument, but can be found in an untriggered burst search.
We use data from the $12$ NaI detectors, whose energy range of $8\,\mathrm{keV}$ to $4\,\mathrm{MeV}$ is sufficient to cover most of the burst emission. Since SGR bursts rarely exhibit radiation above $200\,\mathrm{keV}$, we restrict ourselves 
here to photons with energies in the range of $8$--$200\,\mathrm{keV}$.
Additionally, we only used detectors with viewing angles to the source $< 60^{\circ}$, and checked whether the source was occulted by the spacecraft and the other instrument on board \fermi\, the Large Area Telescope (LAT)\@. 

We use the combined samples of bursts from \citet{vonkienlin2012} and \citet{vanderhorst2012}, and include a total of $367$ bursts with available TTE data in our analysis. The duration of a burst, $T_{90}$, is defined as the time in which the central $90\%$ of the photons, starting at $5\%$ and ending at $95\%$, reach the detector.  We extract data from this time interval and add $20\%$ of the $T_{90}$ duration on either side of start and end time to ensure the entire burst is within our data set. Due to their complex structure, defining exactly which temporal features in the light curve belong to a single burst, and which should be regarded as separate bursts, is 
not entirely unambiguous. For this sample, \citet{vanderhorst2012} \fermi/GBM searched data with a time resolution of $0.064\,\mathrm{s}$. Consecutive time bins in excess of $5.5\sigma$ 
(in the brightest detector; $4.5\sigma$ for the second-brightest detector) were defined to constitute individual bursts. In addition, for two events to qualify as separate bursts, the time between their time bins with the maximum count rate in the TTE data had to be greater than a quarter of the neutron star's spin period ($0.25\times2.072\,\mathrm{s} \approx 0.5\,\mathrm{s}$) and the count rate had to drop back to background between consecutive bursts. 
The burst identification performed in \citet{vanderhorst2012} was not perfect, but it was consistent. However, we would like to point out here that using their definition, 
some ambiguity remains: if the bursts originate high up in the magnetosphere, there could be a waiting 
time longer than $0.5\,\mathrm{s}$ between individual peaks in a single burst. Conversely, the condition that the count rate in all bins within a burst must
exceed the inferred background count rate makes the exact burst definition instrument-specific, such that a single burst could be regarded as two separate events
if part of the variability is hidden underneath the background level.

Here we will nevertheless proceed with the definitions from \citet{vanderhorst2012} to identify individual bursts, with intra-burst variability to be further characterised using the 
methods described below in the \ref{ch6:methods}. In what follows, we will use the term ``burst'' to refer to a light curve with significant transient X-ray emission from the magnetar, as defined above. In contrast, we will use the term ``spike'' for features {\it within} bursts that we attempt to model with the methods described below.

For every burst, the photon arrival times 
are barycentered before analysis, i.e.\ projected to the centre of mass of the solar system, to account for the effects of the relative motion of the spacecraft and the Earth.
 While we could perform the analysis on the unbinned TTE data, this turns out to be computationally prohibitive for any large sample of bursts. Thus, we bin the data to a
 time resolution of $0.5\,\mathrm{ms}$. This time resolution is chosen small enough to probe short time scales, while at the same time allowing for computational efficiency.
Photon arrival times recorded with \fermi/GBM are affected by both dead time and saturation. 
Dead time occurs because the instrument cannot record a second photon within $2.6\mu\mathrm{s}$ of arrival of a previous photon. 
The second photon is thus either not recorded at all, or, on occasion, recorded as a single photon with the combined energy of the two. These deletions and combinations 
impose a minimum time scale onto the data not predicted by a Poisson model, and thus when time-binning photons, the resulting statistical distribution of the counts in a bin will
deviate from the expected Poisson distribution. This effect scales with count rate, such that higher count rates lead to stronger deviations in the statistical distributions of the data.
Dead time is harder to quantify and account for than saturation. It is possible to correct for the fraction of photons lost due to dead time at a given observed count rate using simulations of the instrument \citep{meegan2009,briggs2010,chaplin2013}. While this correction is useful for analyses that rely on accurate estimate of the total number of photons, it is not useful for timing analyses, because the count rate correction will not remove the deviation of the statistical distributions of the counts in a bin from a Poisson distribution. In practice, there may be a weak systematic effect whereby the peaks in the data with the highest count rates $>10^{5} \,\mathrm{counts}\, \mathrm{s}^{-1}$ may be systematically lower by $\leq10\%$ than the actual observed count rate \citep{bhat2014}. 
We currently do not take dead time into account in our analysis. Saturation may also significantly alter the shape of the arriving bursts. Saturation occurs when the number of arriving photons per second measured in a single GBM detector exceeds the maximum data throughput rate of the science data bus. In this case, transmitted rates are capped at a 
count rate of $3.5 \times 10^{5} \, \mathrm{photons} \; \mathrm{s}^{-1}$; any photons exceeding that number are lost. For very bright bursts, this leads
to flat-topped spikes truncated at the maximum count rate. Any bursts with count rates this high are excluded from the sample, as the model we
consider is not designed to represent these features, and continue our analysis with a sample of $332$ unsaturated bursts.

Because \fermi/GBM is wide-field instrument, background from other hard X-ray sources in the field of view is a significant component in the light curves. Additionally, \fermi/GBM serves as a trigger to the larger instrument onboard, the Large Area Telescope (LAT) and will slew to point LAT into the direction of a bright source if triggered by the onboard software. This leads to a changing background on the order of tens of seconds. However, because magnetar bursts are $\sim 2$ orders of magnitude shorter, we can ignore this changing background for the purpose of our analysis. More importantly, the background adds a roughly constant (on the time scale of a magnetar burst) component of $\sim 300 \,\mathrm{counts}\,\mathrm{s}^{-1}$ to each burst light curve. Thus, there may be significant variability in the bursts hidden below this background which we cannot realistically infer. This limitation can only be overcome with a pointed instrument with a much smaller field of view.

\section{Analysis Methods}
\label{ch6:methods}
In order to successfully model the complex temporal variability in magnetar bursts, any modelling procedure must satisfy the following criteria: (1) it must be flexible enough to be applicable to a large number of bursts with distinctly different morphologies. We achieve this by decomposing magnetar burst light curves into one or more components with simple shapes, which, taken together, make up a burst. (2) The procedure must be largely automated, and be capable of inferring both the number of components as well as the model parameters for each component without human intervention. The latter is achieved by setting up a hierarchical statistical model, where the number of components is a parameter to be inferred together with the corresponding parameters of the individual components. We use MCMC (in the form of Diffusive Nested Sampling, or DNS, as implemented in DNest \citep{brewer2011}) to sample the posterior distribution over all parameters including the number of components. From samples of the posterior distribution,
we can then study the properties of individual burst components, as well as their ensemble properties for a given burst.

\subsection{A generative model for light curves of fast transients}
\label{ch6:model}

For a light curve with $K$ bins with Poisson-distributed counts $\bm{\counts} = \{\counts_k\}$, we define a model as a superposition of $N$ individual components:

\begin{eqnarray}
\mean_k &=& \mean_{\bg} + \sum_{n=1}^N \mean_{nk}
 \label{eqn:countsmodel}\\
\mean_{nk} &\equiv& A_n\,\word\left(\frac{t_k+\Delta t/2-t_n}{\tau_n}\right) \; ,
\end{eqnarray}
where $\mean_{nk}$ is the mean count rate of the $n^{\mathrm{th}}$ model component in time bin $k$, 
$\mean_{bg}$ is the background count rate of that bin,
and the count rate $\mean_{nk}$ in a bin $k$ with width $\Delta t$ is defined as the value of a functional form defining the shape of
the model component $\word$ at the mid-point of that time bin. The component $\word$ is a generic shape,
and can be modified by an amplitude parameter $A_n$ and a parameter setting the width $\tau_n$, in addition to
parameters such as the time offset $t_n$ and intrinsic parameters further describing the component's shape.
We will define a component model $\word$ for magnetar bursts in Section \ref{ch6:wordmodel} below, and
restrict ourselves here to a general description of the model.

The posterior probability distribution for all model parameters is given by:
\begin{eqnarray}
p(N, \bm{\alpha},\{\bm{\theta}_n \} \given \bm{\counts}, H) &   \\\nonumber
 						 = & \frac{p(\bm{\counts} \given N, \bm{\alpha}, \{\bm{\theta}_n \}, H)\, p(N, \bm{\alpha}, \{\bm{\theta}_n \} \given H)}{p(\bm{\counts} \given H)} \, .
\end{eqnarray}

Here, $N$ is the number of model components, with the corresponding set of model parameters for these components $\{\bm{\theta}_n\} = \{ \bm{\theta}_1, \bm{\theta}_2, ..., \bm{\theta}_N \}$. Each $\bm{\theta}_n$ may be a scalar, for component models with a single parameter, or a vector, for component models with multiple parameters. 
We build a hierarchical model, where we infer the posterior distributions of parameters $\{\bm{\theta}_n\}$ at the same time as the posterior distributions of hyperparameters $\bm{\alpha}$ describing the shape
of the conditional priors for  $\{\bm{\theta}_n\}$, $p(\{\bm{\theta}_n\}\given \alpha, N, H)$.
$H$ encodes the prior choices we have
made about the model that are not part of the inference problem at this stage: for example, the choice of shape for prior distributions and the model shape used to represent the light curve.

We use a Poisson likelihood to describe the data,

\begin{eqnarray}
\label{eqn:poissonlikelihood}
\likelihood(N, \bm{\alpha}, \{\bm{\theta}_n \}) & = & p(\bm{\counts} \given N, \bm{\alpha}, \{\bm{\theta}_n \}, H) \\ \nonumber
 &= & \prod\limits_{k=1}^{K}{ \frac{e^{-\lambda}\, \lambda^{y_k} }{y_k! }} \; ,
\end{eqnarray}

which only depends on the parameters of the individual model components, $\{\bm{\theta}_n\}$. In general, the conditional priors for the model 
parameters $\{\bm{\theta}_n\}$ will depend on the priors for the hyperparameters defining their prior distributions, $\bm{\alpha}$, such that the
overall prior for this model is

\begin{equation}
p(N, \bm{\alpha}, \{\bm{\theta}_n \} \given H) = p(N | H)\,p(\bm{\alpha}\given N, H)\,p(\{\bm{\theta}_n\}\given \bm{\alpha}, N, H)  .
\end{equation}
Under the assumption that $\bm{\alpha}$ and $N$ are independent, and that the conditional priors of the individual model components are
independent and identically distributed, we can re-write this equation as:

\begin{equation}
p(N, \bm{\alpha}, \{\bm{\theta}_n \} \given H) = p(N|H)\, p(\bm{\alpha}|H)\, \prod\limits_{n=1}^{N}  p(\bm{\theta}_n\given \bm{\alpha}, H) \; .
\end{equation}

Thus, the posterior probability distribution becomes

\begin{eqnarray}
p(N, \bm{\alpha}, \{\bm{\theta}_n \}  \given \bm{\counts}, H) & \\\nonumber
= &  \frac{\prod\limits_{k=1}^{K}{ \frac{e^{-\lambda} \lambda^{y_k} }{y_k! }}\, p(N|H)\, p(\bm{\alpha}|H) \prod\limits_{n=1}^{N}  p(\bm{\theta}_n\given \bm{\alpha})}{p(\bm{\counts} | H)} 
\end{eqnarray}

where the marginal likelihood, $p(\bm{\counts} | H)$ is defined as a normalisation constant in the usual way: 

\begin{eqnarray}
p(\bm{\counts} \given H) & = & \sum_{n=1}^{N} \int_{-\infty}^{\infty}{\likelihood(N, \bm{\alpha}, \{\bm{\theta}_n \})} \times \\ \nonumber
&& p(N\given H)\, p(\bm{\alpha} \given H) \prod\limits_{n=1}^{N}  p(\bm{\theta}_n\given \bm{\alpha}, H) \dd\bm{\alpha}\, \dd\bm{\theta}_1 \, ...\, \dd\bm{\theta}_N \; .
\label{eqn:marginal}
\end{eqnarray}

The marginal likelihood involves integration in a high-dimensional parameter space as well as summation over $N$, which is analytically intractable for all but the
simplest cases. Here, we use DNS to efficiently traverse parameter space and approximate the marginal 
likelihood as well as sample from the posterior probability distribution. 

 \begin{figure}[h]
\begin{center}
\includegraphics[width=9cm]{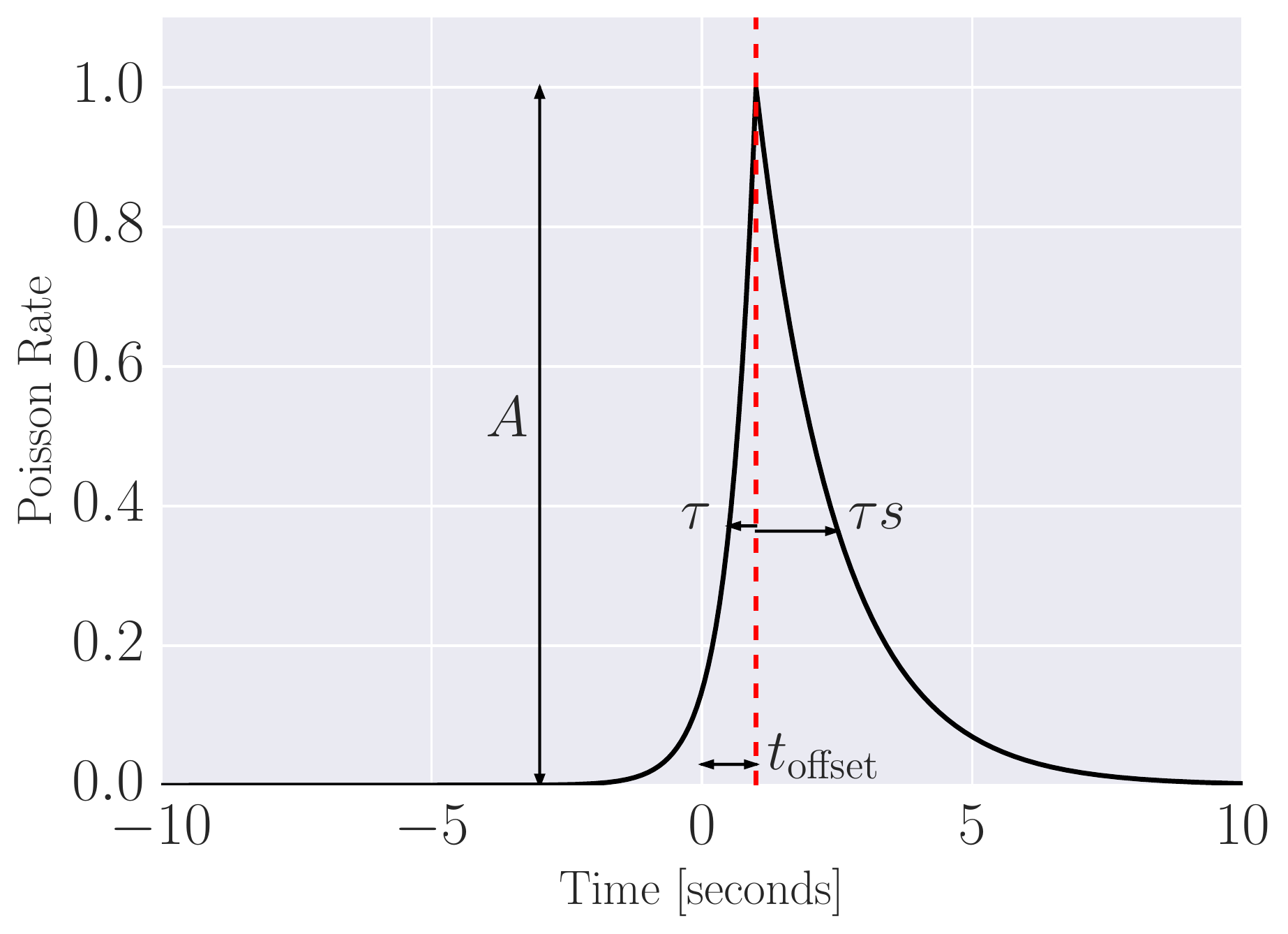}
\caption{An example of the component model used for magnetar bursts: a spike is defined by an exponential rise with characteristic
time scale $\tau$ and an exponential fall with a fall time scale $\tau s$, where $s$ is a skewness parameter that describes how the fall
time is stretched ($s > 1$) or contracted ($s < 1$) compared to the rise time. Individual spikes are also characterised by a time offset
$t_{\mathrm{offset}}$ describing the location of the peak count rate in a time series (the difference between the dashed line and the 0 point
on the x-axis), and an amplitude $A$ describing the height of a peak.}
\label{fig:word_example}
\end{center}
\end{figure}

Nested sampling \citep{skilling2006} samples parameter space by uniformly sampling $M$ points (particles) from the space allowed by the prior. 
One then computes the likelihood values associated with each particle, and the particle with the lowest likelihood is discarded. A new point
can be generated, for example, via MCMC\footnote{In our case, the MCMC process will include trans-dimensional jumps that change the number of components $N$.}, subject to the hard constraint that the
likelihood of the new point must be larger than the likelihood of the discarded one. In this way, the population iterates progressively towards
regions of higher likelihood.

The MCMC step is the key challenge in any Nested Sampling algorithm. Standard MCMC-based Nested Sampling often fails for probability distributions that
are multi-modal. It is here that DNS provides a reliable alternative \citep[for details, see][]{brewer2011}. Instead of discarding all but the last point in the Markov chain, the likelihoods of all parameter sets visited during the MCMC exploration step
 are recorded. One can then compute the $1-e^{-1}$ quantile, and record this value as the new likelihood contour, such that the prior space
 under consideration contracts by a factor of $\sim\!\! e$. Where traditional Nested Sampling now enforces a hard likelihood constraint, Diffusive Nested Sampling samples from a mixture of the 
 two regions, and the particle may escape into the lower-likelihood region, where it can explore more freely. Once enough samples are accumulated,
 one again computes the $1-e^{-1}$ quantile, and constructs a new likelihood contour. This process is repeated, and as in the classical Nested Sampling
 approach, the population contracts towards regions of high likelihood, but it will do so even for complex probability distributions subject to multi-modality.
 Details of an implementation of DNS for hierarchical models with an unknown number of model components can be found in \citet{brewer2013} and \citet{brewer2014}.

DNS as implemented in {\it DNest3}\footnote{the code is available under the GNU public license at \url{https://github.com/eggplantbren/DNest3}} can return both an estimate for the marginal likelihood,
suitable for model comparison, as well as samples drawn from the posterior distribution for a given burst. The latter is done by letting a particle explore
the final mixture of levels freely via MCMC\@.

\subsection{Model shapes}
\label{ch6:wordmodel}

The model defined in Section \ref{ch6:model} is fairly general: it is valid for any Poisson-distributed light curve thought to be composed of several individual components of
the same shape, but with potentially different individual parameters (such as amplitudes and widths). 
We have made only three assumptions so far: (1) the likelihood follows a Poisson distribution, (2) the priors for the number of components $N$ and the hyperparameters $\bm{\alpha}$ are independent,
and (3) the priors on the parameters of the model for the individual components are independent of each other and identically distributed. 
Here, we now refine this model for magnetar bursts specifically. However, changing the model shape for use with different source classes (e.g.\ GRBs) is straightforward.

Figure \ref{fig:example_bursts} shows examples of magnetar burst light curves, each composed of one or more sharp, spike-like features.
We model each feature with an exponential rise and an exponential decay of the type:

\begin{equation}
\word(t) = A \left\{\begin{array}{ll}\exp((t-t_{\mathrm{offset}})/\tau) & \mbox{for $t < t_{\mathrm{offset}}$}\\ 
\exp(-(t-t_{\mathrm{offset}})/(\tau s)) & \mbox{for $t \geq t_\mathrm{offset}$}\end{array}\right. \, ,
\label{eqn:word}
\end{equation}

where $\word(t_i)$ is the component function depending on time parameter $t$ and a skewness
parameter $s$. By giving each component $n$ in the model a time offset $t_{\mathrm{offset},n}$ (equivalent to the time of the peak), 
an amplitude $A_n$ and an exponential rise timescale $\tau_n$ in addition to the skewness parameter $s_n$, 
these spikes can become a representation of the spikes observed in magnetar burst light curves (see Figure \ref{fig:word_example}
for an example of the model). In what follows, we will always use the word ``burst'' for a collection of photons significantly above the background,
using the definition in Equation \ref{ch6:data}, and refer to a ``spike'' when talking about components modelling variability within a burst. 
\begin{table*}[hbtp]
\renewcommand{\arraystretch}{1.3}
\footnotesize
\caption{Model Parameters and Prior Probability Distributions}
\begin{threeparttable} 
\begin{tabularx}{\textwidth}{p{2.0cm}p{10.0cm}X}
\toprule
\bf{Parameter} & \bf{Meaning} & \bf{Probability Distribution} \\ \midrule
\it{Hyperparameters} && \\ \midrule
$\mu_A$ & Mean of exponential (log-normal) prior distribution for spike amplitude $A$ &  $\log(\mu_A) \sim \textnormal{Cauchy}(0, 1)T(-21, 21)$  \\
($\sigma_A$)\footnotemark[1] & Standard deviation of log-normal prior distribution for spike amplitude $A$ & $\mathrm{Uniform}(0,2)$ \\
$\mu_\tau$ & Mean of exponential (log-normal) prior distribution for spike exponential rise time scale $\tau$& $\mathrm{LogUniform}(10^{-3}T_\mathrm{b}, 10^3{T_\mathrm{b}})$\tnote{\emph{b}}  \\
($\sigma_\tau$)\footnotemark[1]  & Standard deviation of log-normal prior distribution for spike exponential rise timescale $\tau$& $\mathrm{Uniform}(0,2)$\\
$\mu_s$ & Mean of uniform prior distribution for spike skewness $s$ & $\mathrm{Uniform}(-10, 10)$ \\
$\sigma_s$ & Half-width of uniform prior distribution for spike skewness $s$& $\mathrm{Uniform}(0,2)$\\ \midrule
\it{Individual Burst Parameters} && \\ \midrule
$t_0$ & Position of spike peak & $\mathrm{Uniform}(t_{\mathrm{start}}, t_{\mathrm{end}})$ \\
$\tau$ & Exponential rise time scale & truncated $\mathrm{Exponential}(\mu_\tau)$, $\tau>\tau_{\mathrm{min}}$ \\
 && ($\mathrm{LogNormal}(\mu_\tau, \sigma^2_{\tau}))\tnote{\emph{a}} $ \\
$A$ & Amplitude of spike peak in units of counts per bin &$\mathrm{Exponential}(\mu_A)$ \\
 && ($\mathrm{LogNormal}(\mu_A, \sigma^2_{A}))\tnote{\emph{a}} $ \\ 
$s$ & Skewness parameter, such that exponential fall time $\tau_{\mathrm{fall}} = s\tau$ & \\
$\counts_{\mathrm{bg}}$ & Flat sky/instrumental background in counts per bin & $\log(\counts_{\mathrm{bg}}) \sim \textnormal{Cauchy}(0, 1)T(-21, 21)$ \\
$N$ & Number of components (spikes) per burst & $\mathrm{Uniform}(0,100)$  \\\bottomrule
\end{tabularx}
   \begin{tablenotes}
      \item{An overview of the model parameters and hyperparameters with their respective prior probability distributions. For parameters where we have explored an alternative distribution in Section 
\ref{ch6:priortest}, we give parameters and distributions for both priors.}
     \item[\emph{a}]{See Section \ref{ch6:priortest} for a discussion on testing an alternative, log-normal prior for spike amplitude and exponential rise time scale.}
     \item[\emph{b}]{$T_\mathrm{b}$: duration of total burst}
\end{tablenotes}
\end{threeparttable}
\label{tab:priortable}
\end{table*}

For each component $n$ we have a set of free parameters $\{t_n, A_n, \tau_n, s_n \}$; for each model, (a 
linear combination of components) we have a set of free parameters $\{N,\mean_{bg}, \{t_n, A_n, \tau_n, s_n\} \}$.
We set a uniform prior on the number of components $N$, where $N$ may lie between $0$ and $100$. 
The priors for the free parameters of the individual components, given the hyperparameters, are independent and identically distributed, such that priors 
are the same for a given type of parameter between components. 
Because bursts are defined in such a way that the observed count rate must drop back to the background before the
start and the end of each burst, we can simply define the prior on the time offset such that $t_n$ must lie between
the start and end times of the burst light curve, $t_{\mathrm{start}} < t_n < t_\mathrm{end}$. 
For the amplitude $A_n$ and the rise timescale $\tau_n$, we choose exponential priors \citep{skilling1998}:

\begin{eqnarray}
    p(A \given \mu_A) &=& {\textstyle \frac{1}{\mu_A}}e^{-A/\mu_A} \\
    p(\tau \given \mu_{\tau}) &=& {\textstyle \frac{1}{\mu_\tau}}e^{-(\tau - \tau_{\mathrm{min}})/\mu_{\tau}}, ~~\tau>\tau_{\mathrm{min}},
\label{eqn:exponential_prior}
\end{eqnarray}

where $\mu_A$ and $\mu_{\mathrm{\tau}}$ are hyperparameters describing the width of the exponential distribution.
We set a hard limit on the minimum possible rise time scale
$\tau_{\mathrm{min}}$ to be a fraction of the light curve's time
resolution, $\tau_{\mathrm{min}} = \Delta t/3$. 

The prior on the skewness parameter $s_n$ is a uniform distribution with a mean of $\mu_s$ and a half-width of $\sigma_s$, such
that the log-skewness must lie in the range $(\mu_s-\sigma_s) < \log{(s)} < (\mu_s+\sigma_s)$. A definition in terms of mean and half-width ensures
that rise and fall times can be asymmetric towards earlier and later times with equal probability. 

The prior on the hyperparameter $\mu_{\tau}$ is log-uniform. The prior range for
$\mu_{\tau}$ depends on the length of the data set, such that $\log{(10^{-3}t_{\mathrm{burst}})} < \log{(\mu_{\tau})} < \log{(10^{3}t_\mathrm{burst})}$.
The priors on the natural logs of the hyperparameters $\mu_A$ and $\counts_{\mathrm{bg}}$ follow a truncated Cauchy distribution between $-21$ and $21$. We chose these hyperparameters for the Cauchy distribution because they allow for a much wider range in amplitudes than we would expect (we have little prior knowledge of the brightness of a spike or the background, other than instrumental limits).
While allowing the hyperparameters to vary from $\sim 10^{-9}$ to $10^{9}$, the Cauchy distribution ensures a $\sim 0.5$ chance of $\mu_A$ and $\counts_{\mathrm{bg}}$ being close to unity.

For a summary of all model parameters, their definition and prior distributions, see Table \ref{tab:priortable}.

\begin{figure*}[htbp]
\includegraphics[width=\textwidth]{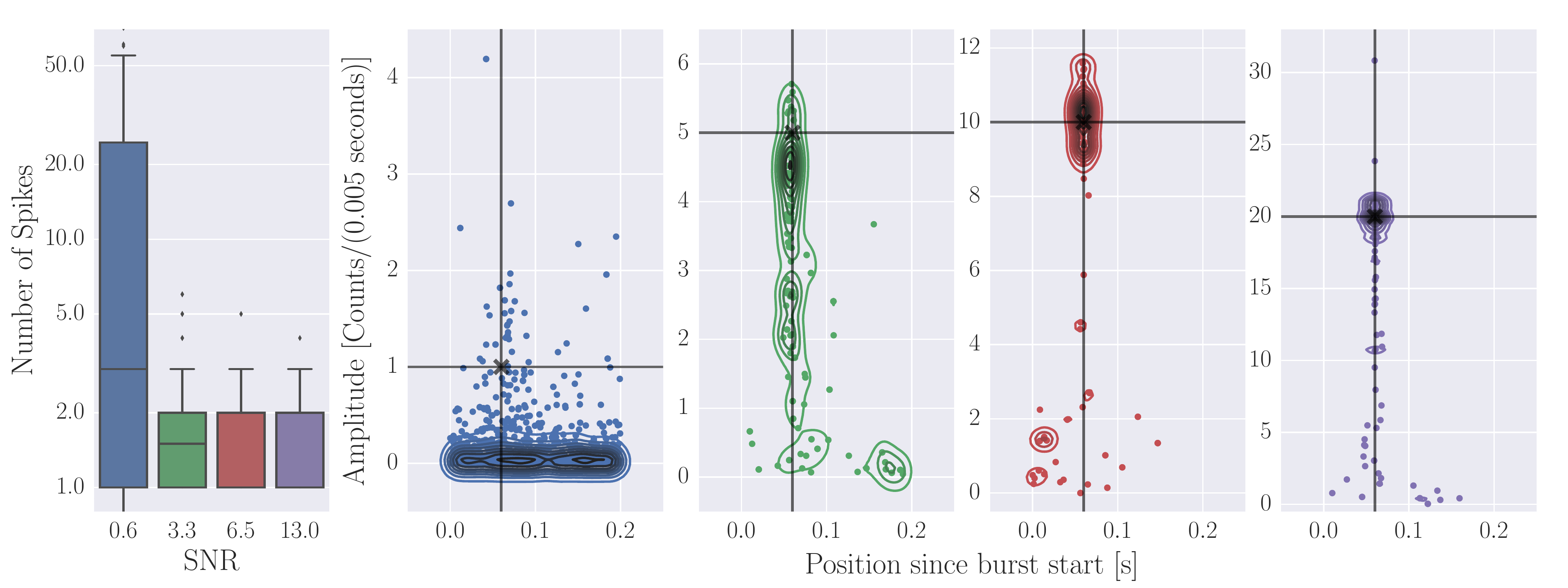}
\caption{We test the model constructed in Section \ref{ch6:methods} on simulated data. We simulated light curves of a single spike with \fermi/GBM-like background count rates and
varied the amplitude of the spike in order to test detectability. In the left panel, we show the posterior distribution of the number of spikes as a function of the signal-to-noise ratio of the spike
 as a box plot. The box encompasses the interquartile range (the $0.25$ and $0.75$ quantiles) with the median marked. The whiskers extend out to $1.5$ times the interquartile range; 
 outliers are marked as scatter points. In the other panels, we show distributions of peak position versus amplitude for the four signal-to-noise ratios of the left panel (in the same order). The position and time and amplitude of the signal injected into the light curve is marked as a dark grey cross; similarly coloured lines are added to guide the eye.} If the noise 
 in the light curve dominates the signal, the model will place a large number of low-amplitude spikes all throughout the light curve (second panel). For a signal-to-noise ratio of 3 or greater, the 
 probability distributions over amplitude and position collapse into a sharp peak at the position and amplitude where we places the spike in the simulations (panels 3-5).
\label{fig:onespike}
\end{figure*}

\begin{figure*}[htbp]
\includegraphics[width=\textwidth]{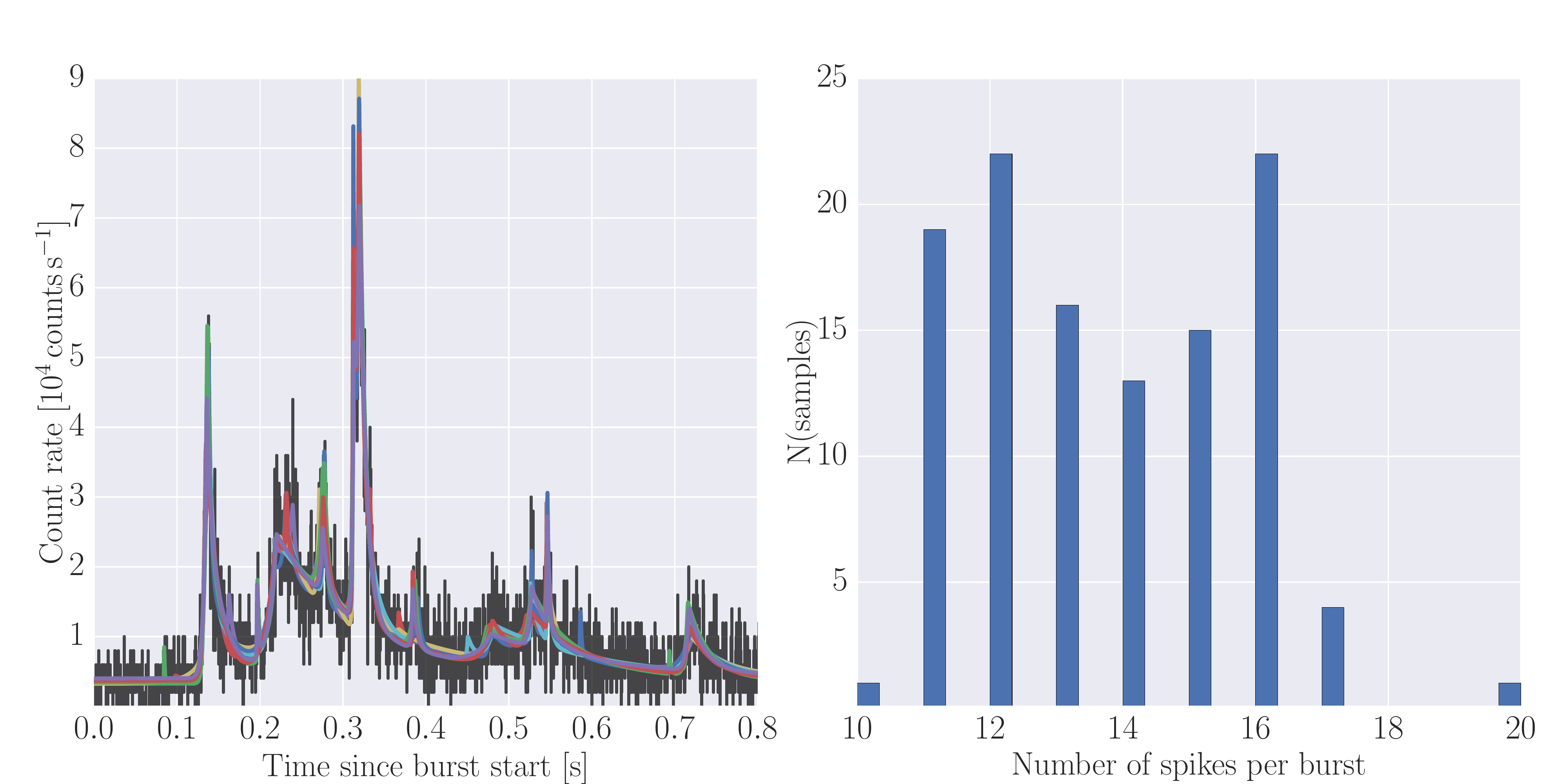}
\caption{An example burst from the magnetar SGR J1550-5418, in an observation taken on 2009 January 22 (ObsID 090122173). In the left
panel, the light curve at high time resolution (black), $\Delta t = 5 \times 10^{-4}\,\mathrm{s}$, and model light curves for $10$ random draws from the posterior distribution (in colours). 
On the right, the marginalised posterior distribution over the number of components in the model. The posterior for the number of components lies between $10$ and $20$ components.}
\label{fig:dnest_example}
\end{figure*}

\section{Simulating from the Model}
\label{ch6:modelsims}

In order to test how well the model and sampling described above work in an ideal case, we simulated burst light curves from the simplest possible version
of the model: a single spike with a constant background. We varied the amplitude of the spike between $1$ and $20$ counts per bin (where $\Delta t = 0.005\,\mathrm{s}$ for
a single bin), but kept all other parameters (duration of the light curve, background noise level, spike rise time scale and skewness) constant. 
We fixed the background count rate to $1.59\,\mathrm{counts}/(0.005\mathrm{s})$, approximated from fitting a flat line to \fermi/GBM background light curves, and the position
of the spike in each light curve at $0.06\mathrm{s}$ from the start of the light curve.

In Figure \ref{fig:onespike}, we show the results for all four simulated bursts. An injected spike with an amplitude below the background level is very hard for the model
to detect, because there is very little information about the presence of spikes in the data. Instead, the spike amplitude will be sampled from a prior with a small inferred 
scale parameter, allowing for a population of spikes with small amplitudes to be present in the data at random positions in the light curve. Most of these spikes will have an 
amplitude much lower than the background itself, reflecting our uncertainty in the presence of any feature given the available data. 

If the burst has an amplitude larger than the background, the model behaves significantly differently. For a signal-to-noise ratio of $\sim 3$ (equivalent to $5$ counts per bin), 
the posterior distribution of the spike position is peaked sharply at the position of the burst peak in the light curve, 
and only few samples are scattered throughout the rest of the light curve. Similarly, the posterior distribution of the amplitude clusters 
around $5 \,\mathrm{counts}/(0.005\mathrm{s})$, with some samples at 
smaller amplitudes largely below the noise level. This scatter can be explained by the uncertainty in the amplitude introduced by the Poisson statistics. 
Additionally, the posterior distribution of the
number of spikes is extremely narrow, with only very few samples indicating more than $3$ spikes in the model. 
Finally, increasing the burst brightness even further leads to an even stronger clustering in amplitude and position, indicating that the brighter spikes are easily detectable with our model 
with little ambiguity. 
Any feature in the data with a count
rate exceeding the background by a factor of at least $3$ will be modelled fairly unambiguously by our method, allowing us to make inferences about the underlying processes based on these models. 

\begin{figure*}[htbp]
\includegraphics[width=\textwidth]{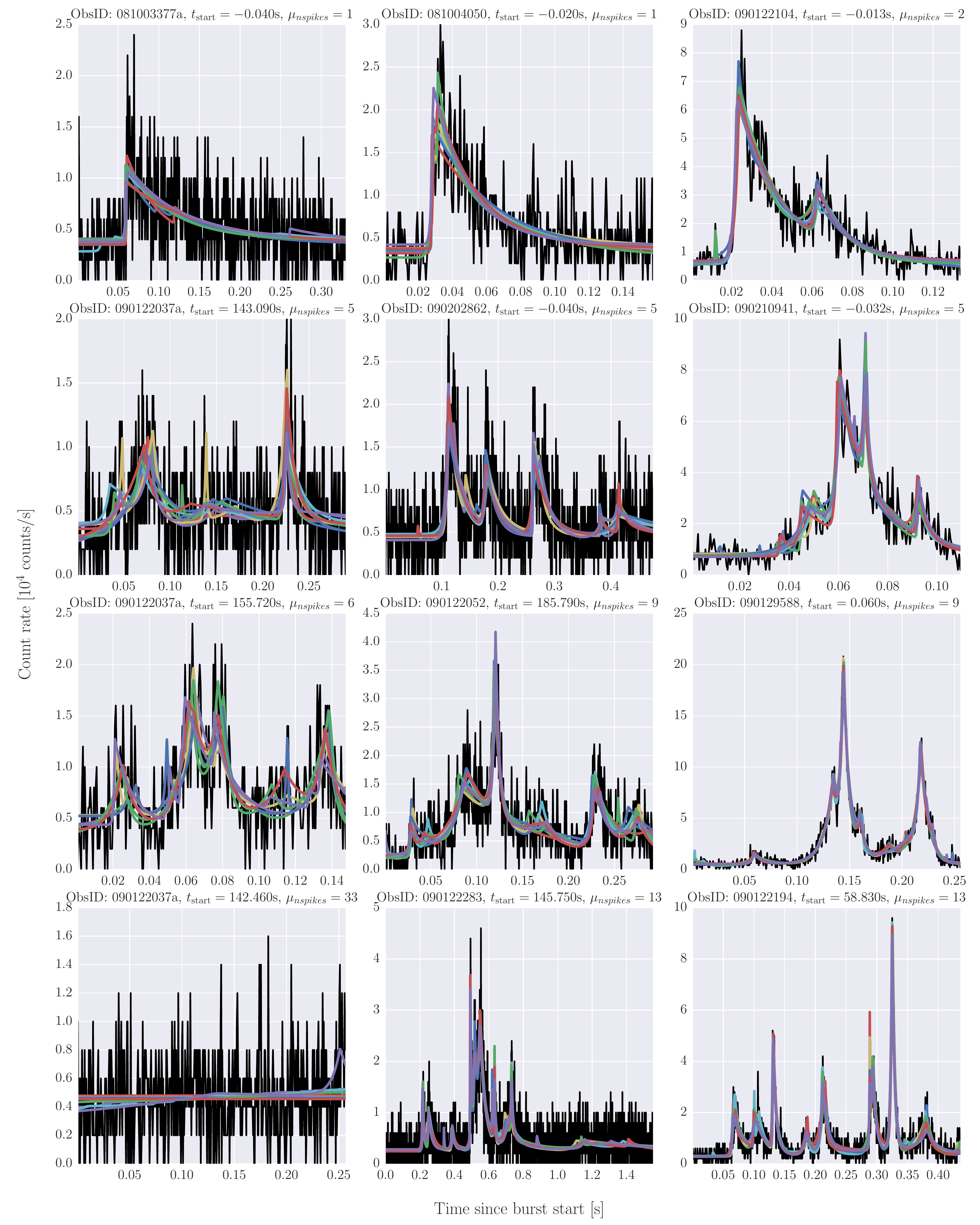}
\caption{Light curves (black) of $15$ examples of bursts from SGR J1550-5418, each with $10$ random draws from the posterior distribution of the model (color). We sorted bursts by maximum count rate 
(left to right) and by the average number of spike components in a given burst inferred from the model (top to bottom). Note that the y-axes are different for each burst, and, in particular, change
quite drastically from left to right. At high count rates 
(with maximum count rates $ > 3\times10^{4}\,\mathrm{counts}\,\mathrm{s}^{-1}$), there are no bursts with an average number of components inferred from the model larger than $15$, and components
 for these models are very well constrained. Conversely, at low count rates $< 3\times10^{4}\,\mathrm{counts}\,\mathrm{s}^{-1}$, there is a population of bursts with a small signal-to-noise ratio, which the 
 model tends to find a large number of low-amplitude spikes at random positions in the light curve (compare last row of panels). This behaviour was also observed in the simulations in Section 
 \ref{ch6:modelsims}.}
\label{fig:lcpanel}
\end{figure*}

\section{Individual Bursts}
\label{ch6:oneburst}

The generative model described above considers each burst individually, and tries to fit the burst light curve with a superposition of spikes with unknown parameters, where the number of
spikes in a burst is not fixed a priori. The results are presented in the form of samples from the posterior distribution of all model parameters for all spikes, the number of spikes,
as well as the hyperparameters listed in Table \ref{tab:priortable}. We can make inferences of individual parameters by marginalising (either integrating for continuous variables, or summing for
discrete variables) over all nuisance parameters (e.g.\ the hyperparameters). 

In Figure \ref{fig:dnest_example}, we show an example of a burst light curve, together with $10$ random draws from the posterior distribution (left panel).
The burst was chosen specifically for its multi-peaked structure such that we can investigate how well the method does in inferring the properties 
of a single burst. Overall, the posterior distribution is narrow and peaked for bright features in the data; the presence of Poisson noise leads to 
uncertainty in the weaker features, leading to a broader posterior distribution in those dimensions.

There is some ambiguity for some features on whether there should be a component, or whether perhaps a
particular feature should be modelled as a superposition of two components, but this ambiguity is generally small. 

If we were interested in deciding whether a feature should be modelled with a spike component, we could marginalise out all other parameters
except for the position parameter and infer the presence of a feature based on its position and the number of samples in which a component appears at
that position. For the quantities we are interested in below, this is not necessary, although the uncertainty in spike position may enter as an 
uncertainty in the distribution of waiting times between spikes.

The number of components (spikes) per burst (Figure \ref{fig:dnest_example}, right panel) has a posterior distribution that is bounded, and all samples drawn
from this distribution contain between $10$ and $20$ components, with most of the posterior mass between $11$ and $16$ components. This range is slightly higher than the features that are immediately obvious to the eye in the burst itself, but may in part be due to some visible features being superpositions of spikes.

In Figure \ref{fig:lcpanel}, we show a grid of light curves with random draws from the posterior distribution of the model for each burst displayed. We choose a wide range
of peak count rates from very weak bursts to the brightest examples in our sample, as well as light curves that vary strongly in complexity, from light curves with only one
 or two peaks to examples with many distinct features. In general, the model will pick out the brightest features and model each of them with a single spike, or a superposition of 
 spikes if the features shows complex structure or a broad peak (the latter are not well matched to our simple exponential model shape). While brighter bursts preferentially have
 models with more model spikes than weaker ones (Figure \ref{fig:lcpanel}, compare left and right columns), the number of spikes for a given burst as well as spike parameters are very well constrained for high signal-to-noise ratios. 
 For the large majority of low-count rate bursts (Figure \ref{fig:lcpanel}, keft column), the model is nevertheless well constrained to few model components per light curve. 
 The sole exception are the very weakest bursts, with very little emission above the sky background.
 In this case, the model has no obvious features to pick out, and instead fits a large number of very-low amplitude spikes at random positions in the light curve (see Figure \ref{fig:lcpanel},
 last row, left panel). \\

\section{Results}
\label{ch6:results}
We explored statistical distributions of model parameters, focussing predominantly on the exponential rise time scale $\tau$, 
the amplitude $A$, and derived quantities such as the fluence (as a proxy for total dissipated energy) and duration for model spikes for
 the whole sample of unsaturated \fermi/GBM bursts from SGR J1550-5418. We sampled from the posterior distribution for each individual burst light curve, then combined
 samples from many bursts to make inferences across the population.
 In a naive way, this can be done by picking a parameter set from the posterior distribution run for each burst, and combining these parameter sets from all bursts to a single ensemble 
to compute the quantity of interest (say, a correlation between two model parameters). This procedure can be repeated, such that we build up a sample
for the quantity in question. 
\begin{figure}[htbp]
\begin{center}
\includegraphics[width=9cm]{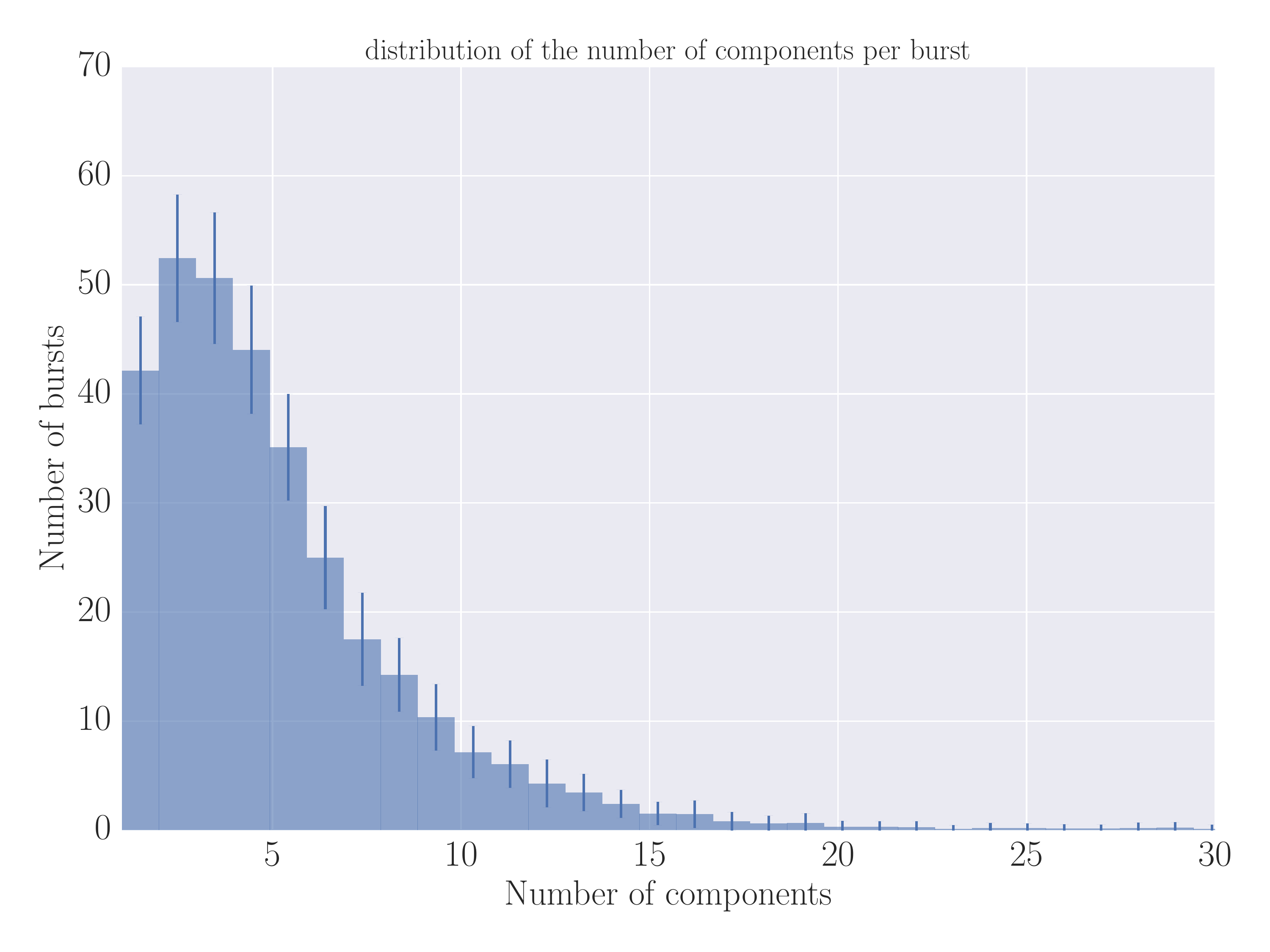}
\caption{Number of components in a given burst for $332$ modelled bursts. We computed the distributions by picking single samples
from the posterior for each burst and formed an ensemble of these individual draws for all bursts to make a single distribution. We repeated this process $100$ times,
and computed the mean number of bursts for each bin (corresponding to the number of components per burst) in the distribution.}
\label{fig:spikes}
\end{center}
\end{figure}

There are some important limitations to this approach. The most important one is the inherent assumption of independence between bursts. By computing and sampling from 
the posterior for each burst individually, we have effectively assumed that each burst is independent of all other bursts; thus each burst has its own posterior distribution for model 
shape parameters as well as hyperparameters. In practice, however, it is more likely that all bursts come from the same underlying physical process, such that we would expect 
bursts to share hyperparameters (i.e.\ the bursts of a given magnetar ``know'' about each other). 
The strong assumption of independence made here can have significant effects on our results: it is, for example, a strong prior against detecting a correlation between parameters. 

Furthermore, we implicitly assume in the following analysis that (1) we model all relevant sources of variability in the data, and (2) that our spike shape model is a reasonable 
representation of spikes in magnetar bursts. If either of these assumptions is broken, the variability in the data not captured by our model may change both the number of 
spikes detected and the shape parameters. 

The last type of uncertainty stems from our prior assumptions. For bursts with a strong signal, the prior is of little importance; for the weakest bursts, however, there is so 
little information in the data that we essentially sample from the prior, which then becomes an important factor. We do a first exploration of the effect our choices of 
priors have on the final conclusions we derive, and defer a detailed analysis across the entire sample of bursts including a model comparison between different priors, 
as well as for a number of different possible spike model shapes, to a follow-up study. We therefore caution the reader that these limitations exist, and results presented 
below should be seen as a first exploratory analysis of the data.

  \begin{figure*}[htbp]
\begin{center}
\includegraphics[width=\textwidth]{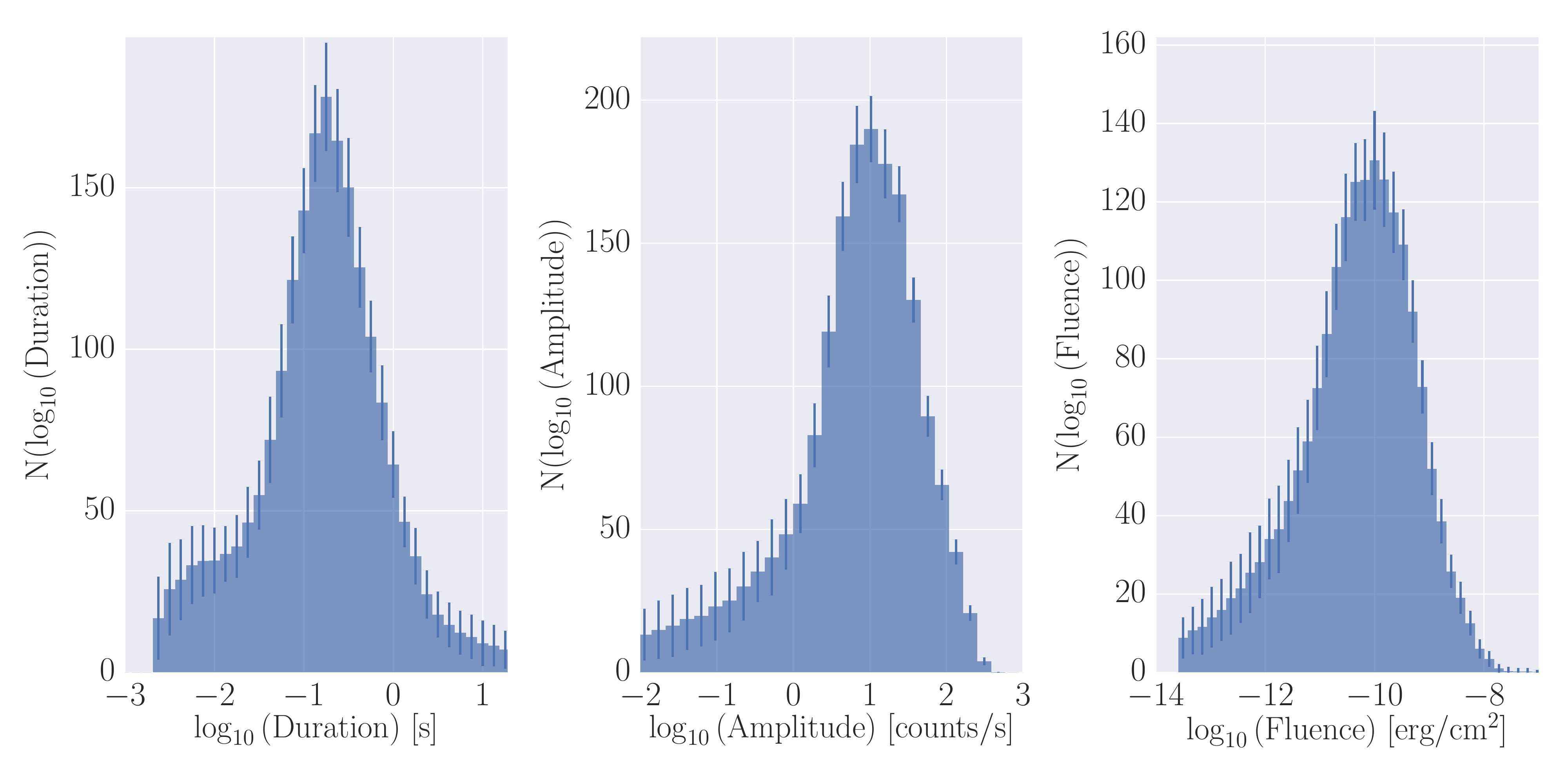}
\caption{Differential distributions for the duration, count-space amplitude and fluence for all model components from $332$ bursts. In each bin, we plot the mean (bars) and standard deviation (error bars)  for
that bin from $100$ ensembles of random draws from the posterior distribution of each burst (see text for details). All three distributions are 
strongly peaked, duration and amplitude seem to have a slight excess at smaller values.}
\label{fig:diffdist}
\end{center}
\end{figure*}

\subsection{Exploring Differential Distributions and Correlations Between Key Quantities}
\label{ch6:exploration}
A first exploration of the data reveals that most bursts can be represented with only a few spikes, of the order of 
$10$ or fewer (see also Figure \ref{fig:spikes}). Because we use very high-resolution data for this analysis in order to
be sensitive to short timescales of $< 1\,\mathrm{ms}$, one source of potential uncertainty in our analysis is the
danger that the model might try to represent individual (e.g.\ background) photons as spike features; this would greatly
skew the resulting distributions towards small amplitudes and very short durations. The fact that we have many bursts
fit with few spikes indicates that this might not be much of a problem. However, we also note that there is an appreciable fraction
of spikes ($\sim 35\%$) with amplitudes smaller than the inferred background count level for the respective model of which these spikes 
are part, similar to what we observe in the simulations presented in Section \ref{ch6:modelsims}. Where the count rate is low in the
data, we cannot say with certainty whether there are (weak) features close in amplitude to the background count rate: they may well exist, but the intrinsic sky and instrumental background
 make it difficult to ascertain their presence. In this limit of low source count rates, the model essentially explores the prior. 
 This problem can be effectively solved with a hierarchical model that considers all bursts at the same time. In this case, all bursts will share the same prior
 distributions with the same hyperparameters. Consequently, in this model weak bursts will be realizations drawn from the tail of those prior distributions, and their parameters will be much better constrained than when seen individually, because all bursts will inform the inference of all other bursts. However, this analysis is beyond the scope of this exploratory work.

Five quantities are of particular interest for their potential connection with physical processes: the spike duration $T$, the exponential rise time scale $\tau$, the exponential fall
time scale as parametrised by the skewness parameter $s$, $\tau_{\mathrm{fall}} = s\tau$, the total dissipated energy $E$ and waiting 
time between consecutive spikes $t_\mathrm{wait}$. The exponential rise time scale and skewness for each model component are free parameters
in our model, and thus easily extracted. We compute spike duration by finding the time between the two points at which the flux has dropped by a
 factor of $100$ on either side of the spike peak. We choose this definition for the spike duration, as opposed to, for example, a definition in terms of where the spike vanishes
 into the instrumental background, because it is independent of the sky background (which may be variable between observations and even between bursts). Thus, by defining the 
 duration in terms of rise time, skewness and amplitude alone, in other words, only in terms of the spike parameters, we avoid introducing unnecessary instrument-dependent
 biases into our analysis.

 Finally, we compute the dissipated energy in a spike by integrating Equation \ref{eqn:word} analytically in count space,
 then converting from count space to fluence using the spectral modelling results from \citet{vanderhorst2012} and \citet{vonkienlin2012}. First, a detector response was generated in order to deconvolve the source spectrum from detector effects. 
 Then, each background-subtracted burst spectrum was fit independently in the $8-200\,\mathrm{keV}$ range. The fluence of each burst was estimated by integrating the 
  energy spectrum estimated by the best-fit spectral model.
 
 We subtract the integrated number of 
 background counts, derived from the background parameter $\mu_\counts$, from the total integrated number of 
 counts in a burst in the full $8$--$200\,\mathrm{keV}$ energy band, and then converted between count space and 
 the fluences computed in \citet{vanderhorst2012}. To compute the fluence in a single spike, we divide the dissipated energy in that spike (integrated analytically)
 by the total number of background-subtracted counts in that burst, and multiply the resulting fraction with the burst fluence. 
   
  \begin{figure}[htbp]
\begin{center}
\includegraphics[width=9cm]{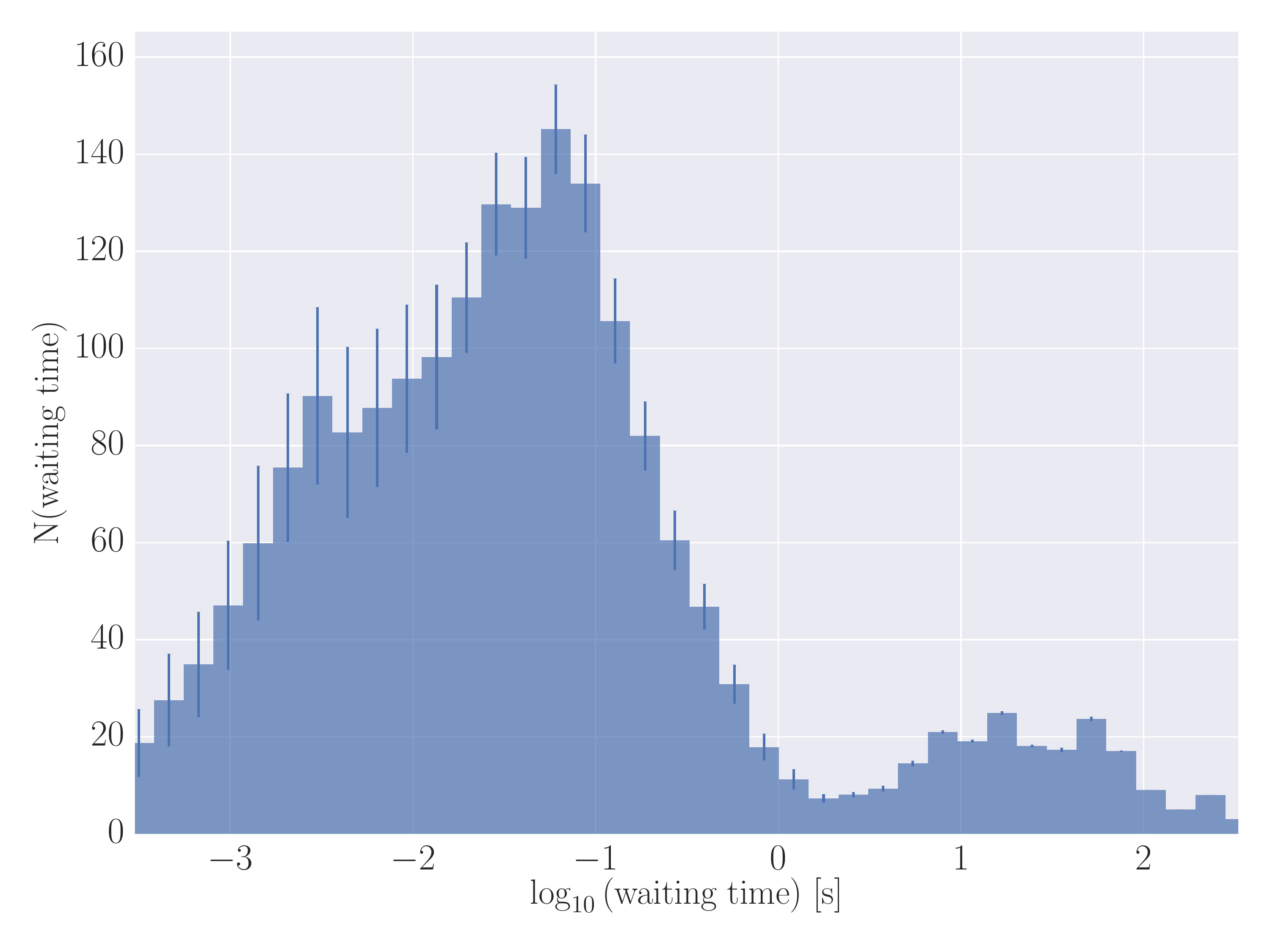}
\caption{Waiting time distribution between consecutive spikes for $332$ bursts. We show the mean (bars) and standard deviation (error bars) of $100$ ensembles of burst models, where each member of the ensemble is a random draw from a burst's posterior. Waiting times are the times between peak amplitudes of spikes, and are only computed for
continuous stretches of data between them, i.e.\ time series without data gaps. This effectively sets an upper limit to the waiting time that can be measured of $330\,\mathrm{s}$, the 
length of a single \fermi/GBM TTE data file. This cut-off introduces a bias into the distribution at long waiting times, and shifts the peak at long waiting times to smaller values.}
\label{fig:waitingtimes}
\end{center}
\end{figure}

\begin{figure*}[htbp]
\begin{center}
\includegraphics[width=\textwidth]{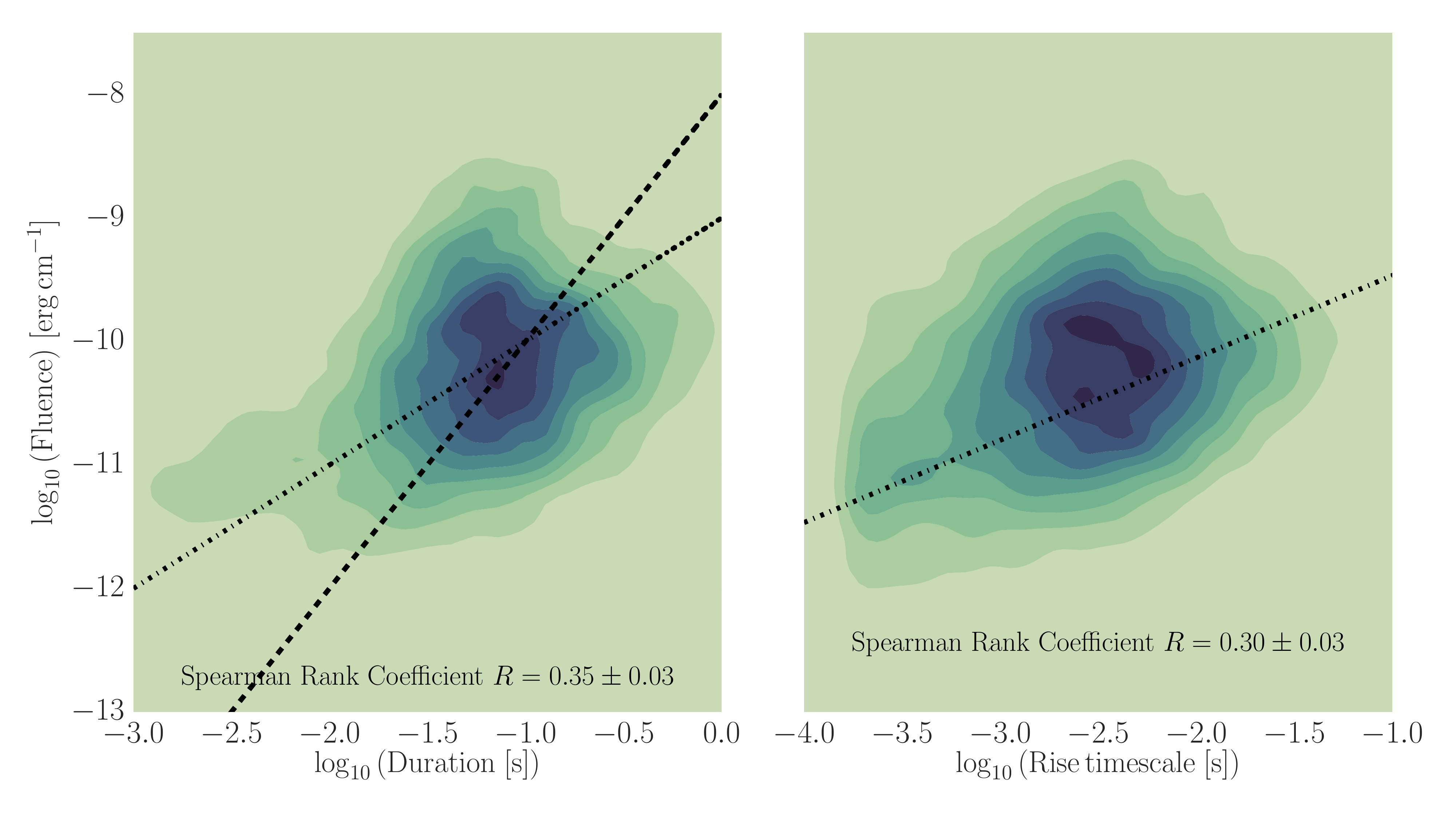}
\caption{Joint distributions of spike duration and fluence (left) and spike rise time and fluence (right). Shaded contours are derived from a single realization of 
draws for all bursts. The duration is defined as the time between the two points where
the count rate drops to $0.01$ of the peak count rate on either side of that peak, computed from the inferred model parameters of each spike. The rise time is the exponential
rise time scale defined in Equation \ref{eqn:word}. The fluence is calculated by computing the fraction of photons within a single spike compared to the number of photons in the
entire burst, and using
this fraction in combination with fluences inferred from spectral modelling \citep{vanderhorst2012,vonkienlin2012} to derive the fluence in a single spike. The dashed line in the left panel corresponds to 
the prediction of the crust rupture model, the dash-dotted line in both panel to the predictions of a reconnection model with a tearing mode instability. Both models are described in more detail in Section \ref{ch6:discussion}.}
\label{fig:correlations}
\end{center}
\end{figure*}
 Here, we test for correlation between fluence and rise time as well as fluence and duration, and construct differential distributions
 in order to compare with predictions from SOC theory. We construct differential distributions by picking a sample from the posterior distribution
 for each burst in the data set, and hence form an ensemble of posterior draws for all bursts, for which we can construct the differential distribution.
 We repeat this process $S$ times, such that we have $S$ ensembles of burst models and $S$ differential distributions, and plot the mean and standard deviation of the distribution in each bin of that distribution. 
 Similarly, we test for correlations by $S$ ensembles of burst models and test for the presence of a correlation using a Spearman rank
 coefficient for each ensemble. We then report the mean and standard deviation of the distribution of coefficients for $S$ draws, where $S = 100$ in all cases below.

 In Figure \ref{fig:diffdist}, we show differential distributions of duration, peak amplitude (measured in count space) and fluence for all spikes in $332$ bursts. All three distributions are
 strongly peaked, and all three have longer tails towards small values. We will consider the behaviour of
 the differential distributions in detail and compare the observed distributions to predictions from SOC theory in Section \ref{ch6:soc}.

In Figure \ref{fig:waitingtimes}, we show the differential distribution of waiting times for the spikes within $332$ magnetar bursts, as derived from 100 random draws for each burst. We use only waiting times 
between consecutive spikes that have no data gaps between them. This effectively sets an upper limit to the waiting time distribution of $330\,\mathrm{s}$,
the length of a single \fermi/GBM TTE data file. We impose this limit because we cannot easily measure waiting times longer than this: any data gaps could 
have included bursts, but because we have not observed them, the recorded waiting times between observed bursts will be longer than the true waiting times
between bursts if there were no data gaps. Thus, by imposing this restriction we avoid measuring much longer waiting times than are present in the system.
On the other hand, we cannot measure any waiting times longer than $330\,\mathrm{s}$, the length of a TTE data file, thus the resulting distribution we measure is
highly skewed towards shorter waiting times.
 The distribution is clearly bimodal: a broad peak at long waiting times has a maximum at $\sim\!\! 20 \,\mathrm{s}$, a
second peak at smaller waiting times, $t_\mathrm{wait} \approx 0.05 \,\mathrm{s}$. This implies that there are two significantly different time scales
in the system, as expected. In our definition, bursts are clusters of individual peaks separated by
waiting times much longer than their durations. It is the time between these bursts that sets the peak in long waiting times. This distribution has a peak about
an order of magnitude lower than that reported for the strongest-field magnetars, SGR 1900+14 \citep{gogus1999} and SGR 1806-20 \citep{gogus2000} as observed
with \rxte\, however, we note that as discussed above, the intrinsic short duration of \fermi/GBM TTE data files we used in this study imposes an 
artificial limit of $330\,\mathrm{s}$ on the waiting times we measure,
and thus introduces a significant skew towards short waiting times. 
The distribution of short waiting times, on the other hand, traces the time between consecutive spikes within a burst. The peak of this distribution indicates
that there is a characteristic time scale determining the waiting time between individual peaks within a burst as well.

\begin{figure}[htbp]
\begin{center}
\includegraphics[width=9cm]{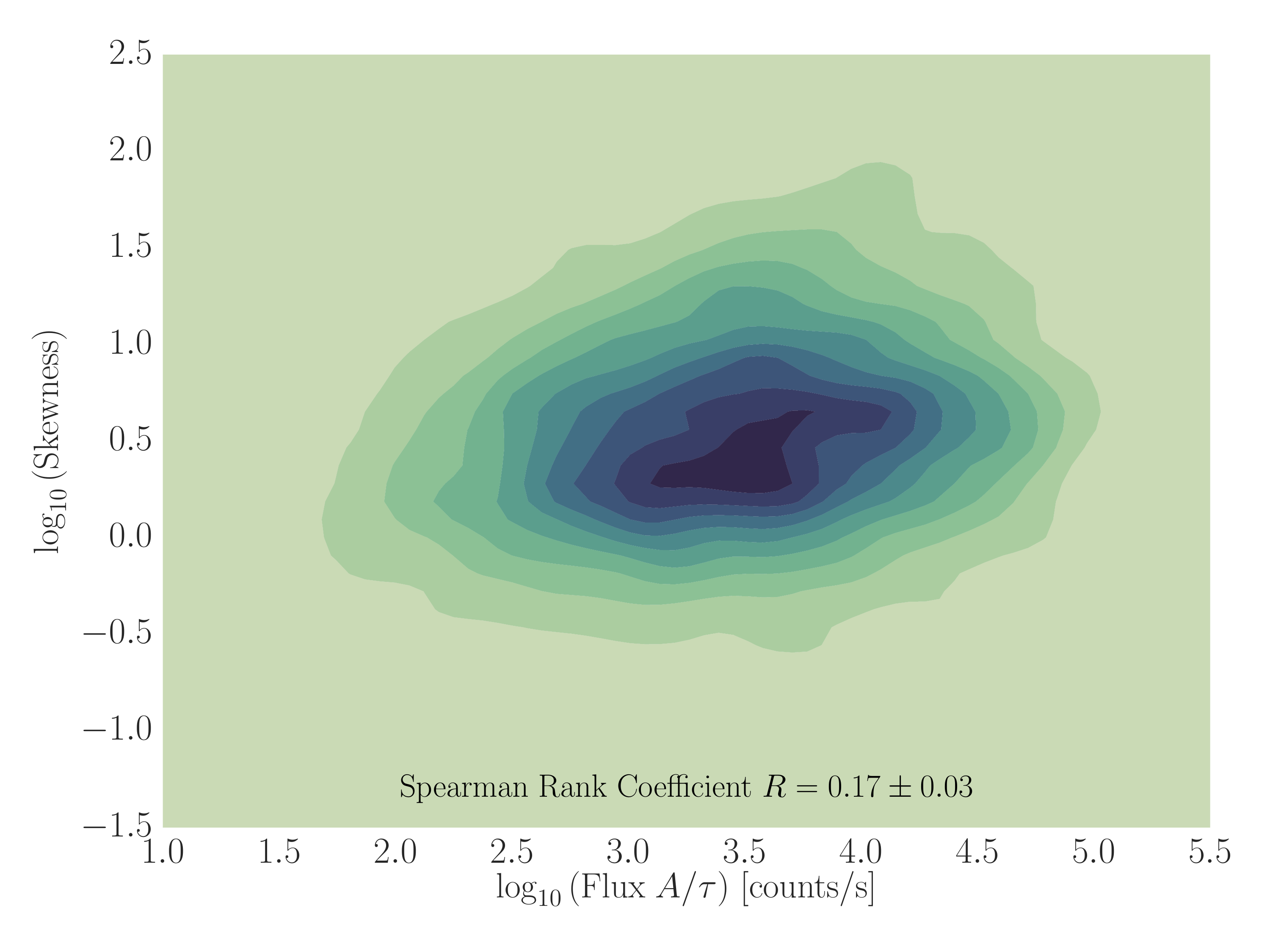}
\caption{Flux versus skewness for a sample of spikes in $332$ bursts. The contours show an ensemble of single draws from the posterior of each burst. We parametrise flux in arbitrary units as the ratio of amplitude $A$ 
for a spike in count space divided by the exponential rise time scale $\tau$. The skewness $s$ measures the asymmetry
of exponential rise versus exponential fall in a spike, such that the fall time scale $\tau_{\mathrm{fall}} = s\tau$. Overall, spikes are skewed towards shorter rise than fall times ($\log_{10}{(s)} > 0$). There is a possible correlation between our measure of the flux and the skewness $s$, with 
a Spearman rank coefficient $R = 0.17 \pm 0.03$ with $p < 3.4 \times 10^{-4}$.}
\label{fig:skewness}
\end{center}
\end{figure}

In Figure \ref{fig:correlations}, we plot rise time and duration against fluence for all $332$ bursts. There is a positive
correlation between both rise time and fluence as well as spike duration and fluence (Spearman rank coefficient $R = 0.35\pm0.03$ with $p < 10^{-5}$ for duration versus fluence,
and $R = 0.30\pm0.03$ with $p < 0.01$ for fluence versus rise time). The small standard deviation between ensembles of random draws from the posterior indicates that the scatter in the correlation depends very little on uncertainties in the modelling. Clearly, more energetic spikes require more time to rise to the peak,
and radiate away their energy for a longer overall time. This is consistent with the results of \citet{vanderhorst2012}, who find a positive correlation between burst duration ($T_{90}$)
and burst fluence, although the correlation found when considering whole bursts instead of spikes seems to be weaker than that found here. It is also consistent with 
previous results in \citet{gogus1999}, who found a correlation between
duration and fluence for bursts from SGR 1900+14 observed with \rxte\, which could be modelled with a power law with an index of $1.13$. 

In Figure \ref{fig:skewness}, we plot a measure of the flux against the skewness parameter $s_n$, in order to test whether bursts with higher luminosity are also more skewed, i.e.\ their ratio of rise time scale to fall time scale deviates significantly from $1$. We parametrize the flux as the peak amplitude $A_n$ of a component $n$ divided by the rise time $\tau_n$ of that component; 
this quantity is effectively a measure of the energy injected into the emission region. If a fireball is formed during the energy release in brighter events, one might expect to see longer decay 
times than otherwise detected, as the fireball would radiate away energy over a longer time interval than if no fireball was formed. On the other hand, fireball formation has no direct 
impact on the rise time, which therefore should remain unaffected. We would therefore expect peaks of larger luminosity to be skewed towards preferentially longer decay times. Overall, the spikes modelling magnetar bursts are skewed \citep[see also ][]{gogus1999,vanderhorst2012}: the population of 
skewness parameters resides largely above 1 (corresponding to $\log_{10}{(s)} = 0$), indicating that most
spikes have faster rise time scales than decays. Moreover, we find that there is a positive correlation between our flux measure and the skewness ($ p < 3.4 \times— 10^{-ˆ'4}$): brighter bursts have a higher ratio between rise and fall time scales. However, due to the large scatter in the distribution, it is not possible to say whether this correlation deviates from a linear relationship 
(as expected for a fireball formation) in the limit of large peak flux.

 \begin{figure}[htbp]
\begin{center}
\includegraphics[width=9cm]{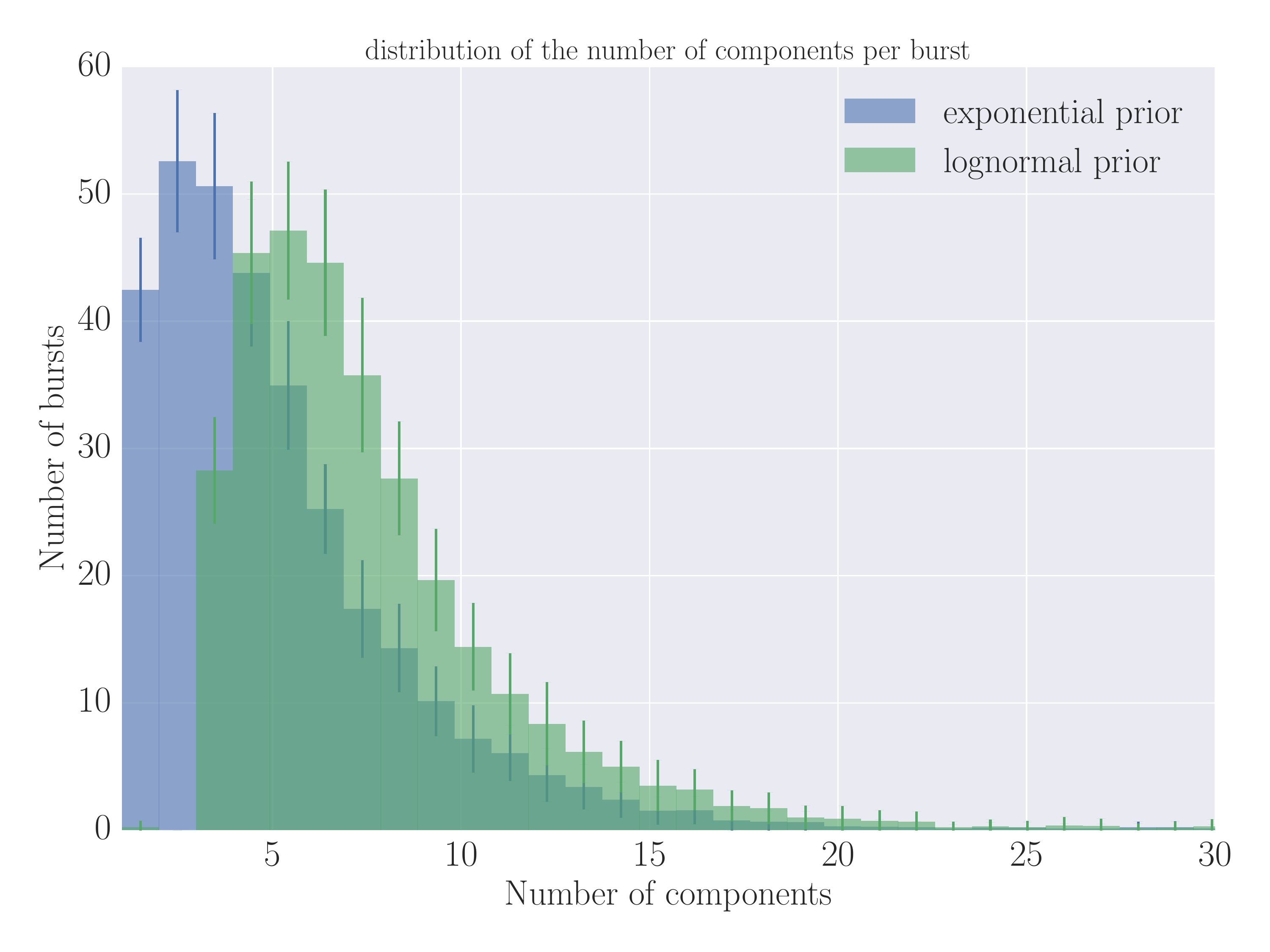}
\caption{The number of  components (spikes) per burst for $332$ bursts for both exponential priors (blue) for amplitude and rise time scale, the standard prior assumed in
deriving the results of this work, and log-normal priors (green) as an alternative hypothesis for the prior distributions on amplitude and rise time scale. Changing the prior to a log-normal
distribution leads to an increase in the number of components per burst.}
\label{fig:nspikes_prior}
\end{center}
\end{figure}
\begin{figure*}[htbp]
\begin{center}
\includegraphics[width=\textwidth]{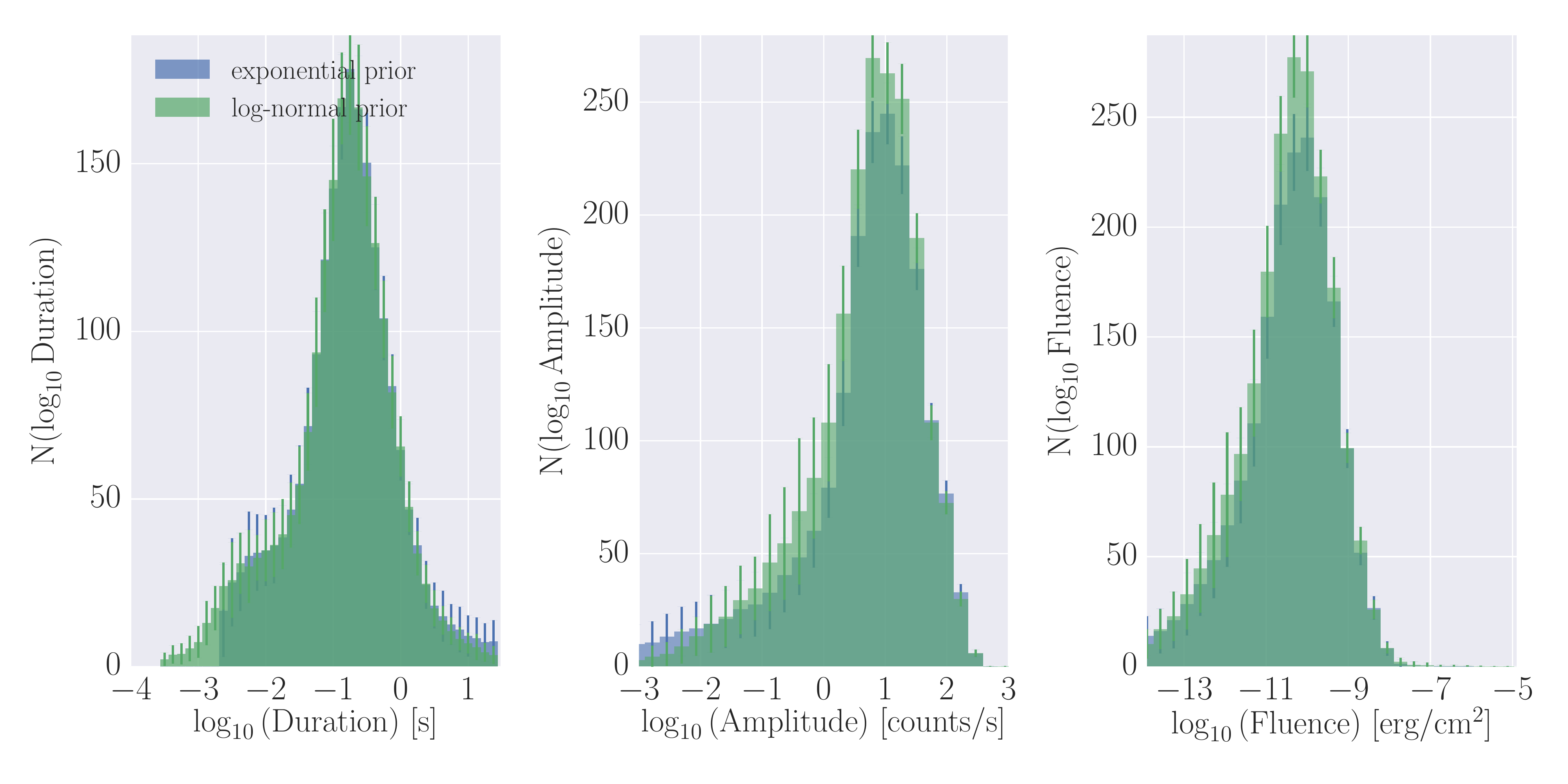}
\caption{Differential distributions for burst duration (left panel), amplitude (middle panel) and fluence (right panel) for an exponential prior on spike 
amplitudes and rise timescales (blue; see also Figure \ref{fig:diffdist}),
and a log-normal prior on the same parameters (green). For each distribution, we drew $100$ realizations from the posterior for each burst and combined them to $100$ ensembles of burst models, for which we computed mean (bars) and standard deviation (error bars).  A log-normal prior leads to a larger number of spikes overall (as shown in Figure \ref{fig:nspikes_prior}), but distributions of
similar shape which are largely consistent within error bars. There is an excess of short-duration spikes when using a log-normal prior compared to an exponential prior. Note also the 
sharp cut-off in duration when using the exponential prior: this is set by the hard limit on the minimum rise time scale, which we did not include in the lognormal prior. 
The reason for omitting this limit in the latter case is the fact that the lognormal distribution, by design, can move its mass away from small time scales, if the data 
demands it, while the exponential distribution cannot. The overall shape and location of the maximum for each distribution, remains unchanged.}
\label{fig:diff_prior}
\end{center}
\end{figure*}

\subsection{Testing the Prior Assumptions}
\label{ch6:priortest}
Because we have little prior information about the nature of magnetar bursts and the details of
the underlying physical processes producing them, we choose mostly uninformative priors for all model parameters. There remains 
some leeway in the choice of these prior distributions, and thus it merits investigation whether changing
any priors has an appreciable impact on the conclusions we derive. 

An important source of uncertainty in terms of deriving conclusions from the data set comes from the priors for
amplitudes and rise time. An exponential distribution favours low-amplitude spikes with short rise time, which, in principle,
could give rise to a population of very short, very small spikes modelling individual photons in high-resolution light curves. We have already shown in 
Section \ref{ch6:exploration} that the generally low number of components per burst disfavours this interpretation, but that there is a population of spikes
 with very small amplitudes. In order
to improve our understanding of how this prior could influence our results, we implemented a log-normal prior for both rise time
and amplitude, with the mean and standard deviations of each distribution free hyperparameters with truncated Cauchy priors on the location parameter of
the log-normal distribution, and a uniform prior on the standard deviations (see also Table \ref{tab:priortable}). We do not perform formal model comparison
(e.g.\ using the marginal likelihood introduced in Equation \ref{eqn:marginal}), but simply explore how the use of different priors
changes the results reported in the previous section.

In Figures \ref{fig:nspikes_prior} and \ref{fig:diff_prior} we present a comparison between the distributions for the number of spikes 
per burst for each set of priors, and the differential distributions for the most important quantities for both priors.
Overall, there is a significant increase in the number of components per burst when using a log-normal prior, related to
an excess in short-duration spikes for a log-normal prior in the differential 
distribution for duration, while the distributions for amplitude and fluence remain qualitatively the same (Figure \ref{fig:diff_prior}) and have an overall excess 
in peaks at the mode of both distributions compared to what is observed in the exponential prior. Much like the differential distributions shown in Figure \ref{fig:diff_prior},  
both the waiting time distribution presented in Figure \ref{fig:waitingtimes} as well as the correlations between various parameters 
presented in Figures \ref{fig:correlations} and \ref{fig:skewness} remain statistically compatible within bounds of the intra-model variance
(comparisons of priors for these choices not shown here), allowing us to remain confident 
in our results in the face of different prior choices.
Why the log-normal prior leads to a larger number of model components is not immediately 
obvious and merits further exploration. However, given that the minor changes in distributions do little to change our conclusions from the data, we defer this 
in-depth exploration of priors to future work.

We also tested whether using binned data rather than the full unbinned Poisson likelihood would have any effect on the derived results. For $50$ randomly chosen 
bursts, we used the same prior assumptions and the same generative model, but an unbinned Poisson likelihood of the form

\begin{eqnarray}
\likelihood(N, \bm{\alpha}, \{\bm{\theta}_n \}) & = & p({t_n} \given N, \bm{\alpha}, \{\bm{\theta}_n \}, H) \\ \nonumber
 &= & \exp(-\int \lambda(t) dt) \prod_n \lambda(t_n) \, \; ,
\end{eqnarray}

where $t_n$ is the nth photon arrival time and $\lambda(t_n)$ is the model flux as defined in Equation \ref{eqn:countsmodel}. The posterior distributions derived with either the binned or the unbinned likelihood are very similar; there is no qualitative difference in the results between the full unbinned likelihood and the binned likelihood implemented in Equation \ref{eqn:poissonlikelihood} .

\section{Discussion}
\label{ch6:discussion}

In this paper, we have introduced a novel way to model magnetar bursts as a superposition of small, spike-like individual components with the same
underlying simple functional form. We successfully decompose a large sample of magnetar bursts from SGR J1550-5418 into individual spikes, and 
find that for the most part, both the parameters of each individual spike as well as the distribution of the number of spikes in a burst are well-constrained.
We then use the inferred marginalised posterior distributions over individual spike parameters to draw first conclusions from the whole sample. 
The differential distributions of exponential rise time scale, amplitude of a spike in count space and fluence are strongly peaked, and skewed with slightly longer
tails at the small ends of the distributions. The differential distribution of the waiting time between consecutive spikes is bimodal, with a peak at short waiting times
of $\sim\!\!0.05\,\mathrm{s}$ and a second peak at $\sim\!\! 20\,\mathrm{s}$. There is a bias at long waiting times due to the observing mode of the instrument, cutting off
waiting times at $330\,\mathrm{s}$. 
We find a strong positive correlation between fluence of a spike and both its exponential rise time scale and its duration. The fluence-duration correlation has been noted
before for bursts as a whole \citep{gogus1999,gogus2000,vanderhorst2012} as a typical signature of self-organised criticality (SOC), which we will discuss in more 
detail below. 
Additionally, there is a correlation between
the inferred flux (defined as amplitude divided by rise time scale) and the skewness in a burst (defined as the ratio of rise time scale to fall time scale). 

There are some important limitations on the results presented above. Most importantly, deriving results over a sample of bursts from posterior distributions for individual bursts
implies a strong prior against detecting a correlation. A formally correct analysis would include the entire sample of bursts in a hierarchical model, where we can explicitly put
priors on relationships between parameters of the model. This is something to be explored in the future.

Secondly, there could be an energy dependence of the results presented above that we have not explored here. Magnetar bursts are known to have spectra that change with flux 
\citep[e.g.][]{younes2013}; conversely, some of the correlations found in this work may be energy-dependent. Again, this would require a more complex model, which is beyond the 
scope of this current work.

Below, we discuss the results of the previous section first in the context of the SOC framework, then compare the derived correlations with predictions of toy models of burst trigger and
emission mechanisms.

\subsection{Comparison with SOC Predictions}
\label{ch6:soc}

Magnetar bursts have often been placed in the context of self-organised criticality. The standard cellular automaton model of \citet{bak1987} models an 
SOC system as a grid of cells (or nodes), where activation of one node leads to a cascade of activation in neighbouring nodes based on some rule 
(e.g.\ a preferred direction). An avalanche is triggered when the critical threshold is exceeded in a single cell. The release in energy can then trigger an event in a neighbouring cell. Depending
on the specific model of the system, triggering any of the neighbouring cells is equally likely, or one may impose that a particular (or several) cell(s) are more likely to be activated than others (e.g.\
if directional fields play a role).
The entire process leads to a sudden energy release and dissipation and return of the system to the sub-critical state.
Based on the avalanche evolution, it becomes possible to define a characteristic length scale and affected 
area (or volume) in the system for that avalanche, which is closely related to the released energy. 

In general, SOC processes follow a fractal geometry, an idea that was proposed by early proponents of SOC theory \citep{bak1989}, and later
confirmed with detailed SOC simulations \citep{aschwanden2012a}, where next-neighbour interactions were found to build up tree-like fractal structures. In the simplest case,
the fractal geometry can be characterised by the Hausdorff fractal dimension $D_d$, which depends on the Euclidean space
dimension $d$. In practice, while it is usually possible to determine the area fractal dimension $D_2$ directly (e.g.\ from solar granulation imaging or solar
magnetograms), this is not true for the $3$D Euclidean space that is relevant for the geometry of both solar flares and magnetar bursts. 
In the absence of firm measurements, it is often reasonable to define a mean fractal dimension, averaging the smallest likely and the maximum possible fractal dimensions.
For most SOC systems, $D_{d,\mathrm{min}} \approx 1$, since for smaller fractal dimensions there would be too little interaction between neighbouring nodes to form an avalanche.
The maximum possible fractal dimension is directly related to the relevant Euclidean dimension $d$. For $d=3$, the mean fractal dimension becomes $D_d \approx (1+d)/2 = 2$. 

For solar flares, there is a wealth of both imaging and timing studies suited to measuring the relevant fractal dimensions. A large body of evidence suggests that area fractal dimension $D_2$ is
close to the expected mean fractal dimension $D_2 \approx 1.5$ \citep[see Table 8 in ][ and references therein]{aschwanden2014}. 
While measurements of $D_3$ are more difficult, \citet{aschwanden2008} found $D_3 = 2.06 \pm 0.48$
in a study of $20$ bright solar flares observed in X-rays, close to the predicted value $D_3 = 2$. 

Simulations of cellular automaton models \citep{aschwanden2012a} and calculations from solar flare observations \citep{aschwanden2012b,aschwanden2013a,aschwanden2013b} 
suggest that the evolution of the avalanche radius $r(t)$ follows a classical diffusion-type relationship after onset of the instability $t_0$, 

\begin{equation}
r(t) = \kappa\,(t- t_0)^{\beta/2}\; ,
\end{equation}

with a diffusion coefficient $\kappa$ and a diffusive spreading coefficient $\beta$.

Using this diffusion law in combination with the fractal dimension $D_d$, the SOC framework makes predictions for 
power law-type correlations between various quantities --- most importantly, duration, 
peak amplitude and total dissipated energy --- as well as their differential distributions.
The predicted correlations for the avalanche duration $T$, peak flux $P$ and total dissipated energy $E$ are

\begin{eqnarray}
P & \propto  T^{d\beta/2} &\propto  T^{3/2} \\ 
E &\propto   T^{D_d\beta/2 + 1} & \propto  T^{2} 
\end{eqnarray}

for typical values $D_d \approx (1+d)/2$, $d = 3$  and  $\beta = 1$ \citep[see ][ and references therein for details of the derivation]{aschwanden2014}. 
For solar flare data in hard X-rays, observed correlations between $T$, $P$ and $E$ are in remarkably good agreement within the measurement
uncertainties with predictions made by the fractal-diffusive SOC model outlined above, with a classical diffusion coefficient $\beta = 1$ and $D_d = 2$.

Similarly, the differential distributions of burst duration, peak amplitude and total dissipated energy can be shown to be

\begin{eqnarray}
\label{eqn:differential_dists}
N(T) dT &=& T^{-\alpha_T}dT \;  \mbox{; $\alpha_T = 1 + (d-1)\beta/2$} = 2\nonumber \\ 
N(P) dA &=& T^{-\alpha_P}dP \; \mbox{; $\alpha_P = 2 - 1/d$} = 5/3 \\
N(E) dE & = &T^{-\alpha_E}dE  \;\mbox{; $\alpha_E = 1 + (d-1)/(D_d + 2/b)$} =  3/2 \; ,  \nonumber
\end{eqnarray}

again for the standard assumptions stated above. Observations of solar flares in hard X-rays with a number of different telescopes
agree with fractal-diffusive SOC predictions; the power law-like distributions in $T$, $P$ and $E$ and the values of $\alpha_T$, $\alpha_P$
and $\alpha_E$ match those of Equations \ref{eqn:differential_dists}.

As expected from SOC predictions, we see a positive correlation between spike duration and 
fluence (a proxy for the total dissipated energy) for spikes in magnetar bursts, similar to that seen when taking the bursts as a whole \citep{gogus1999}. 
The slope of this correlation seems to be flatter than the $T^{2}$ proportionality
expected from classical SOC, although this requires confirmation with a hierarchical model that allows proper inference over many
bursts at the same time.  

 Unlike the correlations between parameters, however, the differential
 distributions for magnetar bursts do not follow the expected power laws: all three quantities show strongly peaked, unimodal distributions.
 While some of that may be an artefact of excluding the weakest, unconstrained peaks (with amplitudes lower than the inferred
 background count rate) from the sample, this selection cannot account for the complete disparity between theoretical 
 expectations and the inferred distributions from the data. In particular the fluence distribution disagrees strongly
 with the derived differential fluence distributions when considering bursts as a whole, which is known to be power law-like over at least 
 four orders of magnitude \citep{gogus1999,gogus2000,prieskorn2012}.
 It is possible that there is a population of weak, low-amplitude peaks even beyond those predicted by the model considered 
 here that we cannot see due to instrumental and sky background.
Alternatively, if there truly is a paucity of peaks at low amplitudes, durations and energies, this would indicate that while bursts as a
whole behave as an SOC system, the driving process that produces the intra-burst variability does not.

The waiting time distribution for magnetar bursts has traditionally been modelled as a log-normal distribution, with limited knowledge 
in the tail on both sides due to two dominant factors. At long waiting times, a lack of continuous monitoring and the resulting data gaps
make an accurate determination of long waiting times problematic. At the short end, there is a fundamental uncertainty in the definition of 
a single burst (see also discussion on burst extraction in Section \ref{ch6:data} and \citealp{vanderhorst2012} for the standard definition of a burst). 
Indeed, \citet{gogus1999}  and \citet{gogus2000} do not model waiting times $<3\,\mathrm{s}$ in order to avoid confusion between 
separate, single-peaked bursts and multi-peaked bursts. 

In the SOC framework, the occurrence of individual avalanches is generally modelled as a random process: the distribution of critical events in time
follows a Poisson process. This process can either be stationary for a constant driving process, or more often the driving process setting the frequency
of burst occurrences is non-stationary in some way. The details of this non-stationarity will determine the shape of the resulting waiting time distribution.
However, even for a non-stationary Poisson process, the resulting waiting time distribution will be a superposition of two or more exponential distributions,
no matter what the time-dependent behaviour of the underlying driving process is. The resulting distribution 
can be approximated as a power law at long waiting times (see \citealp{aschwanden2011} for derivations for various driving processes), which flattens out at 
short waiting times. 

This is not what is observed when decomposing magnetar bursts into individual spike-like features (compare also Figure \ref{fig:waitingtimes}). There is a
clear bimodality that is at odds with the power law-like predictions of SOC theory. Even for a driving process that operates at a low
rate for most of the time, and only occasionally drives a short period of intense avalanching, SOC predicts a power law that flattens at short
waiting times. Again, it may be possible to explain some of the discrepancy with a lack of well-constrained low-amplitude spikes hidden in 
the background, but this does not explain the clear bimodality between long and short waiting times. This may be an indicator that two different
processes, with two different characteristic time scales, operate in producing the bursts and the intra-burst variability, respectively.

\subsection{Comparison with simple models of the burst process}
As outlined in the Introduction, the mechanisms responsible for triggering magnetar bursts, and the associated emission processes,
 are still poorly understood. The basic picture, however, is as follows. Magnetic field evolution in the stellar interior (via ambipolar diffusion,  the Hall effect, and Ohmic decay, see also \citealt{goldreich1992}), leads to the build up of stress in the system. Stress may build up in the solid crust, which resists motion of the 
 magnetic field lines that are frozen into it. Alternatively, if the crust yields plastically \citep{jones2003,levin2012}, stress may build up in the external magnetosphere if the 
 prevailing plasma conditions permit \citep{beloborodov2014}. Bursts occur when this stress is released on a very rapid timescale, be the cause of stress release a crust rupture or a plasma instability. 
 Magneto-hydrodynamic instabilities in the core itself may also play a role.

The stress release process leads to high energy X-ray and gamma-ray emission, which we observe as a magnetar burst. This emission may comprise 
both a non-thermal component (from particle acceleration or scattering), and a thermal component (as parts of the system are heated). The picture is 
further complicated by the fact that energy from the initial stress release event may be `stored' and then released over a longer timescale. Energy may 
be locked up temporarily in the form of seismic or magnetospheric oscillations, or in the form of an optically thick pair-plasma fireball that may get trapped
 within closed magnetic field lines. One of the main goals of magnetar research is to search for unique signatures of these various possibilities, and hence 
 to distinguish between the various mechanisms suggested.

In this paper we have explored the idea that each burst is made up of cascades of stress release (and emission) events. We have done this by fitting each 
burst with sequences of exponentially rising and decaying spikes, and studying at the ensemble properties of those spikes. We now need to examine whether 
the results are consistent with the predictions of the various models. Sadly, detailed model predictions that link up stress release and emission are sparse to 
non-existent, so a fully rigorous comparison is not yet possible. As a prelude to more detailed studies we can nonetheless use simplified models to explore 
how this type of analysis might allow us to distinguish the different mechanisms.

In connecting physics to observables, we need to be aware of the limitations of what we can infer from our measurements. In particular the following 
considerations apply: (i) The fluence that we measure is the energy released in the X-ray/Gamma-ray waveband. As such it is a lower bound, since 
energy may be lost at other wavelengths or even in other forms such as neutrinos (depending on the precise location of the energy release). (ii) The
 rise time that we measure is the rise time of the emission process, not necessarily that of the stress release process. (iii) The duration that we measure 
 may be the duration of the stress release event, or may be set by the prolongation mechanisms discussed above. We will revisit these caveats in the 
 discussion that follows.

\subsubsection{Three Toy Models for the Waiting Time Distribution}

We start by considering the bimodal distribution of wait times (Figure \ref{fig:waitingtimes}). Within the cascade picture, the wait times represent the 
time for information about the change in configuration resulting from a local release of stress to be communicated to the next failure point (be that a 
weakened part of the crust, or another part of the magnetosphere). The distribution is bimodal, with one peak at $\Delta T\sim 0.05\,\mathrm{s}$ that 
is the wait time between individual spikes, and one at longer timescales (of at least $10\,\mathrm{s}$) that is the timescale between individual bursts. 
We would like to understand whether the distributions are consistent with the timescales on which such information might be conveyed either (a) through 
the crust (set by the shear timescale), (b) through the stellar interior, or (c) through the magnetosphere (the latter two are dominated by their respective Alfv\'en timescales).

\begin{figure}
\centering
\subfigure{
                \includegraphics[width=0.45\textwidth]{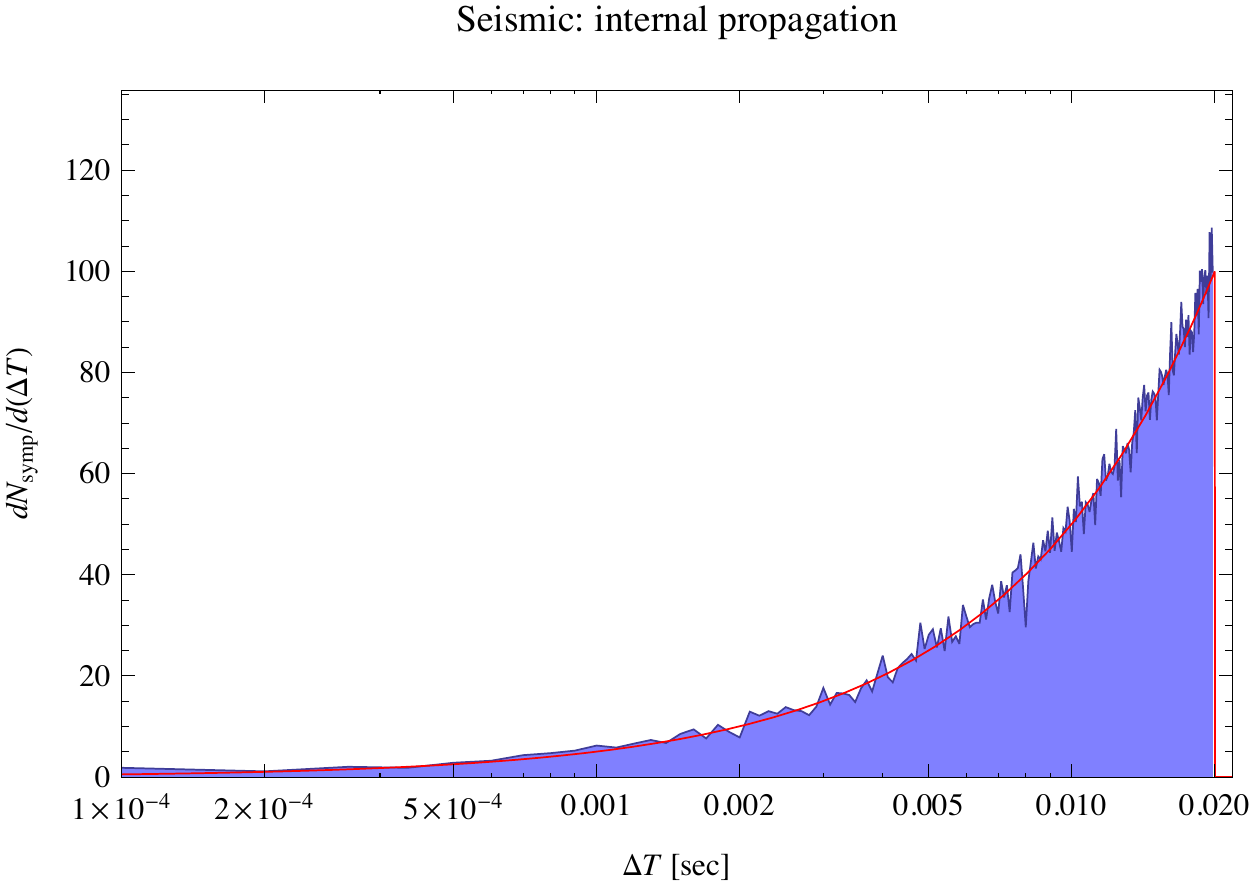}
                \label{fig:sip}}
                \vspace{-0.1cm}
        \subfigure{
                \includegraphics[width=0.45\textwidth]{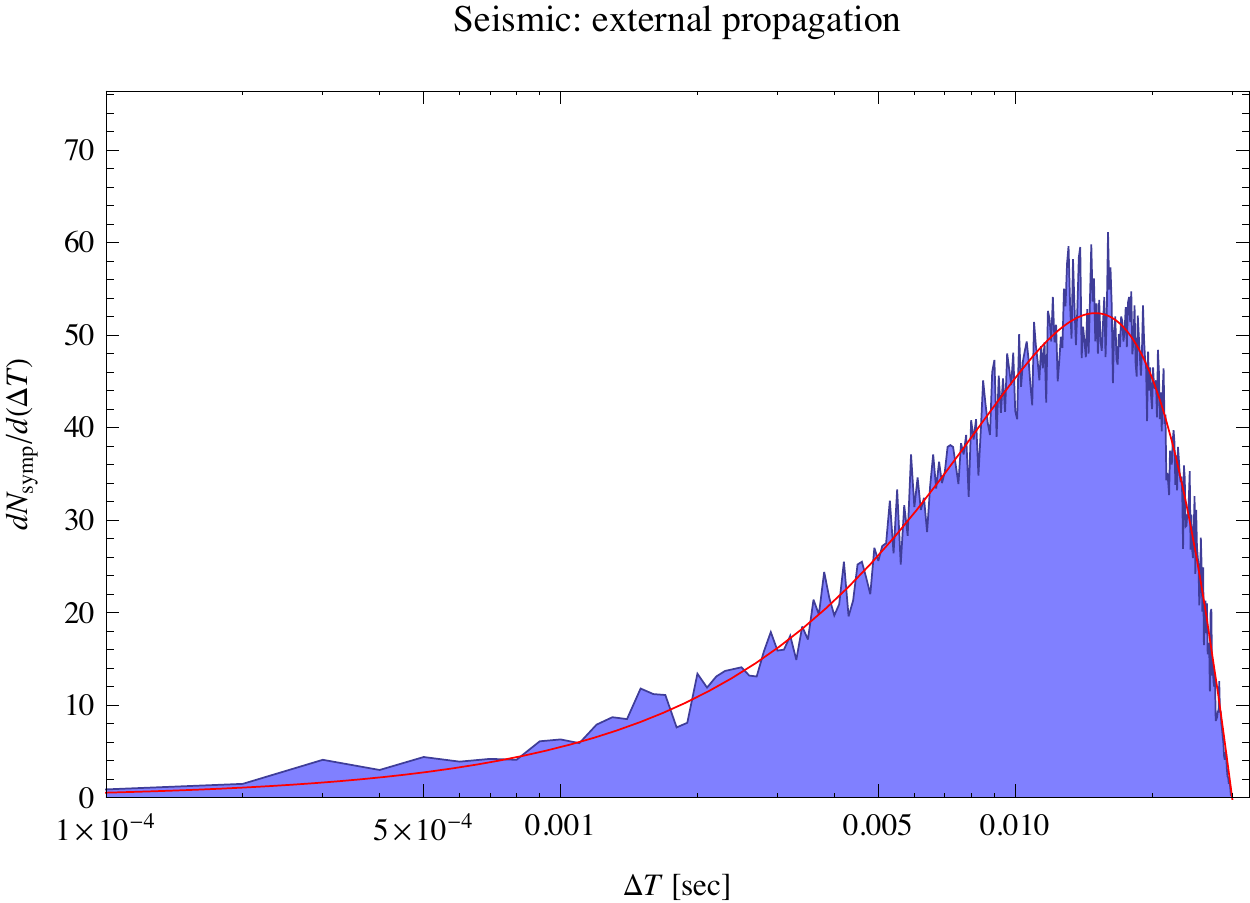}
                \label{fig:sep}}
        \vspace{-0.5cm}
        \subfigure{
                \includegraphics[width=0.45\textwidth]{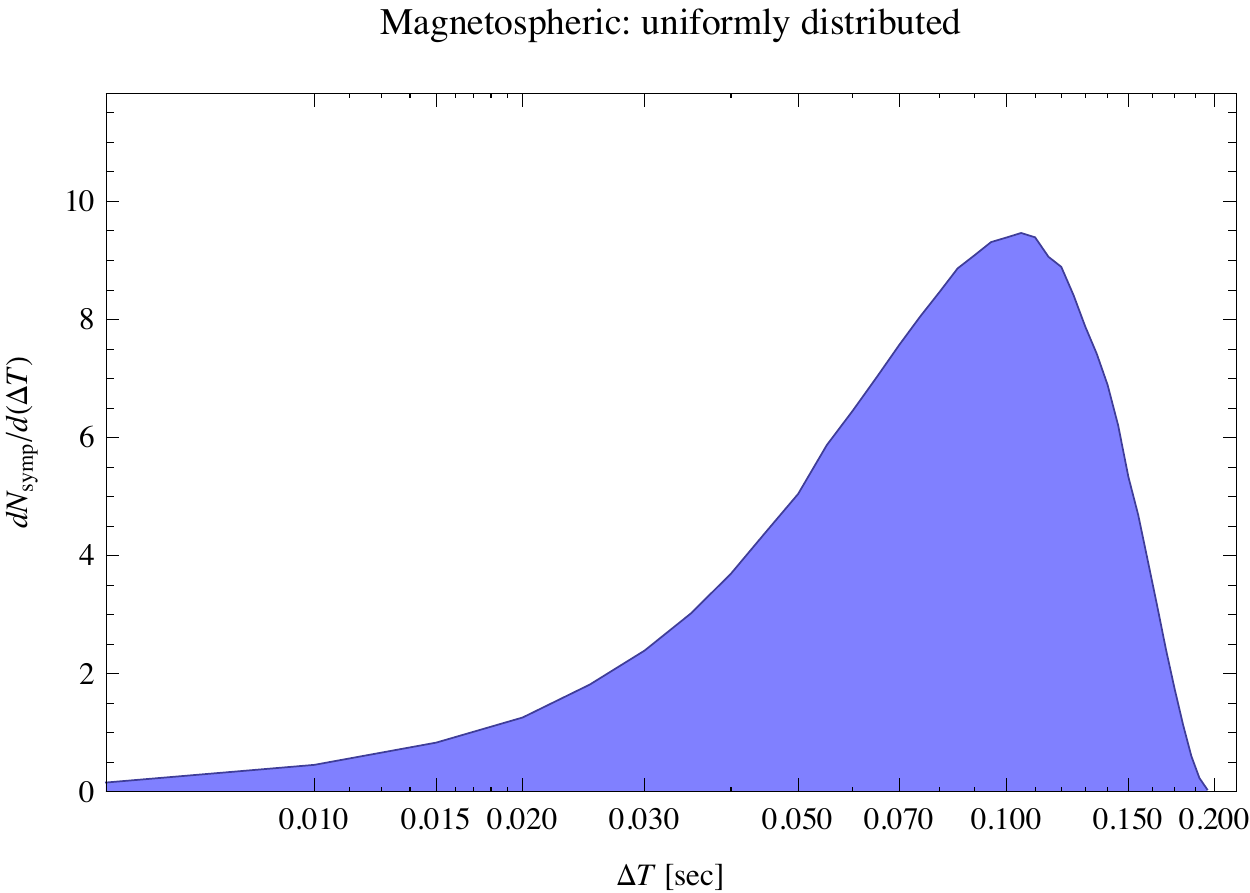}
                \label{fig:mud}}
         \vspace{0.5cm}
        \caption{Waiting time distributions for three different trigger and propagation models. In all cases, we randomly distribute $10^{6}$ points 
        of potential failure (which we also call weak points) 
        either on the sphere of a neutron star of radius $10^{6}\,\mathrm{cm}$ or in the magnetosphere above the star. We then trigger a single point of 
        failure, and compute how much time it would take for that 
        information and energy release to propagate either through the neutron star interior (middle panel), along the surface in the crust (upper panel),  
        or through the magnetosphere (lower panel) to other points of failure. The red line presents the analytic solutions for the first two cases presented
         in the text, the blue distribution the waiting time distribution from $10^{6}$ Monte Carlo-simulated connections between weak points
         constructed as the set of all travel times between failure points for the three different models, as described in the text.}
                \label{fig:wtmodels}
\end{figure}

For transmission of information via Alfv\'en waves through the neutron star interior, the appropriate timescale is set by the Alfv\'en speed 
$v_A = B/\sqrt{4\pi\rho}$ where $B$ is the magnetic field strength, giving

\begin{equation}
v_A = 10^8 \mathrm{cm/s} \left(\frac{B}{10^{16} \mathrm{G}}\right) \left(\frac{10^{15} \mathrm{g/cm}^3}{\rho}\right)^{1/2}.
\end{equation}
This yields timescales for transmission of information through the neutron star core via torsional Alfv\'en waves of $\lesssim 2R/v_A = 0.02 (R/$10 km) s.
 Note however that the value of the field strength $B$ in magnetar cores is highly uncertain, as is the appropriate value of the density $\rho$. In principle 
 only the charged component ($\sim$ 5--10\% of the core mass) should participate in Alfv\'en waves, reducing $\rho$, however there are mechanisms 
 associated with superfluidity and superconductivity that can couple the charged and neutral components \citep{vanhoven2008,andersson2009}. As such this value is also consistent with the 
 inter-spike timescale distribution. In Figure \ref{fig:wtmodels} (top panel panel), we show the expected distribution of waiting times for propagation through 
 the neutron star interior. 

The distance between two random points on a sphere of radius $R_*$ may be written in terms of the colatitude angle $\phi\in[0,\pi]$. Here we adopt 
spherical coordinates ($r,\phi,\theta$) and orientate our coordinate system such that one point is located at ($R_*,0,0$) and the other point at ($R_*,\phi,0$) -- 
note that due to rotational symmetry the distance is independent of the azimuthal angle $\theta\in[0,2\pi)$. The distance between the two points is then
\begin{equation}\label{eq}
s(\phi)=R_*(2-2\cos\phi)^{1/2}.
\end{equation}
Next we choose to write $\phi$ in terms of $x\in(0,1]$,
\begin{equation}\label{eq:phi(x)}
\phi=\arccos(2x-1),
\end{equation}
such that we obtain
\begin{equation}\label{eq}
s(x)=2R_*(1-x)^{1/2}.
\end{equation}
Dividing the above equation by the appropriate velocity, i.e.\ the core Alfv\'en speed  $v_A$, gives us the following transformation function
\begin{equation}\label{eq:trans1}
t(x)=\frac{2R_*}{v_A}(1-x)^{1/2},
\end{equation}
which describes the waiting time $t$ in terms of the variable $x$. Consequently, we assume that the random variable $X$ obeys a uniform density distribution $U(0,1)$:
\begin{equation}\label{eq:uniform dist}
u_X(x) = \left\{
\begin{array}{ll}
1 &  {\rm if~} 0<x\leq1,\\
0 & {\rm otherwise}.
\end{array} \right.
\end{equation}
Then together with Eq.~(\ref{eq:trans1}), we find that the probability density function (PDF) for the variable $T$ is given by
\begin{equation}\label{eq:interior_pdf}
f_T(t) = \left\{
\begin{array}{ll}
\frac{1}{2}\left(v_A/R_*\right)^2t &  {\rm if~} 0<t\leq2R_*/v_A,\\
0 & {\rm otherwise}.
\end{array} \right.
\end{equation}

To derive this distribution, we uniformely distributed a number of weak points across the crust of a neutron star with typical radius following Equation \ref{eq:uniform dist}, and then assumed failure at one 
initial point. We then calculated the time it would take for the information about that failure to propagate through the neutron star interior. It is assumed 
that the impact of the original failure is sufficient to trigger subsequent failure events in other weak points; the waiting time distribution is then the time 
for the information to travel between these weak points.

In Figure \ref{fig:wtmodels} (top panel), we show the expected distribution of waiting times for transmission of information about failure points through the 
neutron star interior. We show both the analytical solution (from Equantion \ref{eq:interior_pdf}) as well as the result of $10^{6}$ Monte Carlo simulations
 of the process.
The expected distribution is clearly at odds with what is observed: while it extends to values high enough to explain the long waiting times, the distribution 
is dominated by a power law with a positive exponent and a sharp drop-off at $\sim\!\! 0.04\,\mathrm{s}$. This is clearly not observed in the data, where 
the inter-spike waiting times resemble much more a log-normal distribution. However, the model assumed here does not include seismic waves traveling 
through the star multiple times. Instead, it is assumed that the energy transferred in the first instance is sufficient to trigger a follow-up event, such that the 
waiting times are directly related to the distance through the star between two points on the surface. The sharp drop-off at long waiting times in this case 
corresponds to the travel time between the furthest two points on the star (i.e. $2R$). It is possible that seismic waves could be reflected and travel through 
the interior multiple times, transferring small amounts of energy every time, and that then the waiting time is determined by the time it takes for the cumulative 
energy transmitted to a weak point to be sufficient to trigger a starquake. However, including these effects would require much more detailed knowledge about 
the energy threshold for crust failure than is available, and beyond the scope of this simple toy model.  

For the second toy model, where information propagates through the neutron star crust only, the governing factor is the shear speed in the 
crust, $v_s = (\mu_s/\rho)^{1/2}$ where $\mu_s$ is the shear modulus and $\rho$ the density. The shear modulus is of the order of the Coulomb potential 
energy $\sim Z^2e^2/r$ per unit volume $r^3$, where $r\sim(\rho/Am_p)^{-1/3}$ is the inter-ion spacing, while $Z$ and $A$ are the effective atomic number 
and mass number, respectively, of the ions in the crust. Using the shear modulus computed by \citet{strohmayer1991} and scaling by typical values for the 
inner crust \citep{douchin2001}, the shear velocity is:

\begin{eqnarray}
v_s & = & 1.1 \times 10^8 \mathrm{cm/s} \left(\frac{\rho}{10^{14} \mathrm{g/cm}^3}\right)^{1/6} \left(\frac{Z}{38}\right) \\
&& \times \left(\frac{302}{A}\right)^{2/3} \left(\frac{1-X_n}{0.25}\right)^{2/3} \nonumber \, ,
\end{eqnarray}
where $X_n$ is the fraction of neutrons (see also \citet{piro2005}). The timescales for transmission of information via shear waves around the crust are 
therefore $\lesssim \pi R/v_s = 0.03$ ($R$/10 km) s. This is certainly consistent the values that we found for the wait times between individual spikes. 

The distance between two points on a sphere of radius $R_*$ that is measured across the surface of the sphere, may equally be represented solely in terms 
of the colatitude angle $\phi\in[0,\pi]$. Following a similar procedure as before we find that the distance between two points, as measured across the surface of 
the sphere, is then
\begin{equation}\label{eq}
s(\phi)=R_*\phi.
\end{equation}
Using Eq.~(\ref{eq:phi(x)}) we get
\begin{equation}\label{eq}
s(x)=R_*\arccos(2x-1),
\end{equation}
which leads to the following transformation function when divided by the corresponding velocity, i.e.\ the shear speed in the crust $v_s$, 
\begin{equation}\label{eq:trans2}
t(x)=\frac{R_*}{v_s}\arccos\left(2x-1\right),
\end{equation}
where the waiting time $t$ is expressed in terms of the variable $x$. Assuming again that the random variable $X$ is distributed according to 
Eq.~(\ref{eq:uniform dist}) and with the transformation function Eq.~(\ref{eq:trans2}) we finally obtain the PDF for the variable $T$ in this case
\begin{equation}\label{eq:crust_pdf}
g_T(t) = \left\{
\begin{array}{ll}
\frac{1}{2}v_s\sin(v_st/R_*)/R_* &  {\rm if~} 0<t<\pi R_*/v_s,\\
0 & {\rm otherwise}.
\end{array} \right.
\end{equation}

Again, we have assumed a number of weak points on the neutron star surface, but this time, propagation proceeds entirely around the star under 
the assumption that transmission proceeds purely through shear waves in the crust, assuming the canonical values given in the expression for shear 
speed above.  As above a single failure triggers an energy release that is communicated through the star and triggers failures at other weak points in 
the crust; the waiting time distribution now depends on the distance between two points on the surface of the star, and the crustal shear speed. 

The distribution of wait times that might result is shown in the middle panel of Figure \ref{fig:wtmodels} for both the analytical solution given in Equation
 \ref{eq:crust_pdf} and $10^{6}$ simulations. In general, this model reproduces the observed waiting time distribution in Figure \ref{fig:waitingtimes} fairly 
 well: it peaks at similar time scales ($\sim\!\! 0.01\,\mathrm{s}$, with a sharp drop at long waiting times and a long tail towards shorter waiting times that 
 is more pronounced than we observe in the data. However, we note that the waiting times may be artificially shortened by line-of-sight effects: if a single 
 event at a point triggers two successive events at different positions on the star or in the magnetosphere, we might see all three in rapid succession 
 depending on our line of sight. This may lead us to conclude that there is a small waiting time between successive (causally connected) events, whereas 
 in reality, two of the three hypothetical events in this example are unconnected.

For the last model, transmission through the magnetosphere, the Alfv\'en speed is of order the speed of light. An upper limit for the waiting time between 
causally connected spikes is given by the light crossing time of the circumference of the light cylinder ($2\pi R_L=2 \pi c/\Omega$) i.e.

\begin{equation}\label{eq}
\Delta t^{\rm max}\sim2\pi \frac{R_L}{c}=P,
\end{equation}
where $\Omega$ is the spin frequency of the neutron star and $P$ the spin period. For magnetars $P\sim2$--$12\,\mathrm{s}$. The typical 
value $\Delta T\sim10^{-2}\,\mathrm{s}$ roughly corresponds to a height of $r_{\rm rec}\sim10^{-2} c/(2\pi)\sim5\times10^7\,\mathrm{cm}$. 
If we distribute reconnection events in the magnetosphere in the region [$r^{\rm min}=10^{-4}c$, $r^{\rm max}=10^{-1}c$] such that the number 
of events per radius remains constant, we find the waiting time distribution shown in Figure~\ref{fig:wtmodels}, lower left. Here, we have distributed 
weak points uniformly in a shell of inner radius $r^{\rm min}$ and outer radius $r^{\rm max}$ around the star, and assumed that propagation of 
energy and information to other weak points must proceed through this shell, or, for simplicity, through the star if the shortest distance between
two failure points intersects the volume occupied by the neutron star.. 
The waiting time distribution expected from the simple toy model is 
in remarkably good qualitative agreement with the observed distribution (compare Figure \ref{fig:waitingtimes}): it peaks at roughly at the right 
waiting time, and has an extended tail towards smaller waiting times. This tail is more pronounced in the toy model compared to the observations, 
but this could potentially be due to observational effects (such as the low-amplitude events that we cannot constrain reliably from the data, and 
which were excluded from this analysis).

We conclude that the shorter `inter-spike' wait time distributions are, given the extreme simplifications of our models, at least in principle 
compatible with transmission of information via crustal shear, magnetospheric, or core Alfv\'en waves.  The latter possibility was most 
problematic in terms of fitting the distribution, however we caveat that our choice of stress points in the magnetosphere (which will have a 
major effect on the predicted wait time distribution) was highly ad hoc.   The longer wait times, by contrast, are clearly inconsistent with any 
of the communication timescales described here, and are therefore more likely to be set by a slower evolutionary timescale of the system.  

\subsubsection{Simple Models of Spike Properties}
We now move on to consider more complex models and to exploit the other spike properties discussed in this analysis. We will explore three 
scenarios: (i) crust rupture trigger that leads directly to high energy emission; (ii) crust rupture trigger that leads to magnetic reconnection, 
with the emission coming as a consequence of that process; and (iii) magnetospheric trigger leading to spontaneous magnetic reconnection 
and associated emission.

We will first explore the crustal rupture model of \citet{thompson1995}. Within this picture, the small magnetar bursts are triggered by crustal 
ruptures as the crust comes under stress from the evolving interior field. The high energy emission that follows comes from particle acceleration, 
scattering and, they argue, a high likelihood of pair plasma fireball formation\footnote{\citet{heyl2005b} also discuss the crust rupture mechanism, 
however in the scenario that they outline, energy is transported by fast MHD waves before forming a pair plasma fireball.}

The energy released in a burst scales in this model \citep[Equation 28ff.\ in ][]{thompson1995} as 

\begin{eqnarray}
\Delta E \sim & 4\times 10^{40} & \left(\frac{L_{\mathrm{thick}}}{10^{4}\,\mathrm{cm}} \right) \left( \frac{L}{10^{5}\,\mathrm{cm}}\right)^2  \\ \nonumber
&& \times \left( \frac{B}{10^{15}\,\mathrm{G}}\right)^{-2} \left(\frac{\theta_\mathrm{max}}{10^{-3}}\right)^2 \, \mathrm{erg} \, ,
\end{eqnarray}

where $B$ is the internal crustal magnetic field, $\theta_{\mathrm{max}}$ is the yield strain of the crust, and $L^2$ is the area of a patch of crust in 
which the energy is released\footnote{Note that the relevant equation in \citet{thompson1995} is missing a length scale to be dimensionally consistent. 
We assume here that this missing length scale could reasonably be the thickness of the crust, $L_\mathrm{thick}$.}. Now if the duration of the observed 
spike $T_\mathrm{dur} \sim L/v_s$ ($T_\mathrm{dur}$ in this section is the theoretical equivalent to the spike duration $T$ measured in Section 
\ref{ch6:exploration}), so that duration is indeed set by the rupture process itself, then we would predict that $\Delta E \sim T_\mathrm{dur}^2$. 
This is qualitatively consistent with the observations (Figure \ref{fig:correlations}): there is a positive correlation between the observed durations and fluences of the spikes, although 
it seems slightly less pronounced than the theoretical calculations here suggest. 

Another interesting question is whether the fluence of a burst always scales in the same way with overall duration; or whether there are deviations 
above a certain energy threshold that might indicate the development of a longer-lived source of emission such as a pair plasma fireball. If this occurs 
then duration would no longer be set by the rupture process, but would instead be controlled by the fireball evaporation timescale.  There is no direct 
evidence in the observed correlations for a change of the correlation with high fluences (Figure \ref{fig:correlations}). Due to the large scatter it is not 
straightforward to see whether the correlation changes at the high end, however, a more complex hierarchical model envisioned for a future work would 
allow us to make a formally valid inference of whether the data can distinguish between these two hypotheses (change in the correlation at high fluences 
versus no change).

The second two scenarios that we explore both involve the high energy emission being generated as a result of magnetic reconnection in the magnetosphere, 
albeit with different original triggers. In general, the total energy output of a spike generated by reconnection ($E$) is given by the total volume that is 
reconnected ($V\simeq L_xL_yL_z$) for the duration of the spike ($T_{\rm dur}$) times the local magnetic energy density $u_{\rm B}=B_S^2(R_*/r)^5/8\pi$, 
where $B_S$ is the surface field, and $r$ is the height of the reconnection region.  For now we consider a twisted external field, for which we use the relation 
$B(r)\sim B_S(R_*/r)^{5/2}$ \citep{thompson2002}.  This yields

\begin{eqnarray}\label{eq:total energy reconnection}
E\simeq V\,u_{\rm B} & \simeq &  L_xL_yL_z\frac{B_S^2(R_*/r)^5}{8\pi} \\ \nonumber
				 & \simeq & L_xL_z\left(T_{\rm dur}\frac{\delta}{\tau_{\rm rec}}\right)\frac{B_S^2(R_*/r)^5}{8\pi},
\end{eqnarray}

where we have used $L_y=T_{\rm dur}v_{\rm rec}=T_{\rm dur}\delta/\tau_{\rm rec}$, with $v_{\rm rec}=\delta/\tau_{\rm rec}$ denoting the reconnection 
speed, $\delta$ the half-thickness of the current sheet and $\tau_{\rm rec}$ the reconnection timescale. The above estimate is not based on any particular 
reconnection mechanism, but merely assumes that the reconnection speed remains constant for the duration of the spike. 

Let us now consider two different scenarios for the reconnection. Consider first of all the case where reconnection is spontaneous, and driven by the 
tearing mode instability, as discussed by \citet{lyutikov2003}. In this case, $\tau_{\rm rec}=\tau_{\rm tm}\propto\delta^{3/2}B_S^{1/4}r^{-7/8}$. Furthermore, 
from the necessary condition for the growth of the perturbation, i.e. $\tau_{\rm tm}<L_x/c$, where $\tau_{\rm tm}=(\tau_R\tau_A)^{1/2}=(\delta/c)\mathcal{S}^{1/2}$ 
(with $\tau_R=\delta^2/\eta$ is the resistive timescale, $\tau_A$ is the Alfv\'en timescale, and 
$\mathcal{S}=\tau_R/\tau_A=\delta c/\eta$ is the Lundquist number), we get $L_x\sim\delta\mathcal{S}^{1/2}\propto\delta^{3/2}$. Note that the 
growth rate of the tearing mode does not depend on the size of the reconnection region in the $z$-direction, i.e. $L_z$. We further assume that the duration 
of the spike is also independent of $L_z$. Combining with Eq.~(\ref{eq:total energy reconnection}) we obtain

\begin{eqnarray}\label{eq}
&E &\propto\tau_{\rm tm}^{2/3}T_{\rm dur}B^{7/4}r^{-33/8}. 
\end{eqnarray}

Thus if the bursts are due to reconnection driven by the tearing mode, fluence should be linearly proportional to the duration of the spike ($E\propto T_{\rm dur}$).  
Assuming that the observed rise time is the reconnection time (as discussed earlier this assumes that emission is essentially instantaneous), this 
yields $E\propto\tau_{\rm rise}^{2/3}$.  This is qualitatively consistent with the observation of a positive correlation between the exponential rise timescale
 and the fluence of a spike (Figure \ref{fig:correlations}, right panel): although the correlation seems to be somewhat steeper than expected from the calculation 
 above, the exact power law exponent is difficult to tell due to the large scatter in the data.   Note that this large scatter in the fluence versus rise time plots 
 may be due to the fact that the energy is also strongly dependent on the local magnetic field strength ($E\propto B_S^{7/4}r^{-33/8}$), where the energy of 
 the spike will be highly dependent on the height of the reconnection region $E\propto r^{-33/8}\sim r^{-4}$. 
 While the scatter in the data makes the determination of the power law exponent difficult, it is not that large from a physical perspective: given the very strong dependence
 of the energy on the height of the reconnection region, a scatter over $\sim 2$ orders of magnitude nevertheless implies that 
the spikes all occur at relatively the same height ($r\sim r_{\rm rec}$), where the local conditions for reconnection are 
favourable, instead of happening all throughout the magnetosphere.

Now consider an alternative scenario, where reconnection is driven directly by crust rupturing, with the speed of reconnection being governed by the 
shear speed in the crust $v_s$.  Assuming again that the observed rise timescale is the reconnection timescale, but in a situation where $\delta$ is 
independent of the reconnection timescale, we would now predict $E \propto T_\mathrm{dur}/\tau_{\rm rise}$.  The dependence on rise time is quite 
different from that predicted for the tearing mode, and is in fact not consistent with the observed correlations.  Although a linear relationship between 
duration and fluence is certainly possible (see Figure \ref{fig:correlations}, left panel), a negative correlation between the rise time and fluence is strongly 
disfavoured (Figure \ref{fig:correlations}, right panel).

The cascade models that we have examined in this section are clearly all highly simplified.  They illustrate, however, the potential diagnostic power of 
the analysis technique presented in this paper, and we hope that the existence of more rigorous techniques like this will motivate the development of 
detailed theoretical models that link trigger and emission mechanisms in a self-consistent manner.  

\section{Conclusions}

In this paper, we have, for the first time, characterised variability within magnetar bursts in detail, using a probabilistic and flexible model for the burst light curves.
The complex temporal morphology of magnetar burst light curves can be very well modelled as a superposition of individual flare-like spikes, represented by a simple
model of an exponential rise and an exponential decay. We use a Bayesian hierarchical model, combined with MCMC sampling, to infer the probability distributions for
parameters of the individual spikes as well as the number of spikes in a burst in an informative and statistically rigorous way. 

We find that light curves are well modelled by varying numbers of spikes, which follow clear trends in their properties. In particular, one can relate model parameters such as the 
rise time and skewness as well as derived quantities such as the duration and fluence to physical quantities in the system. 
While magnetar bursts overall seem to fit within the framework of self-organised criticality, we find that the individual spikes do not: differential distributions of important SOC 
quantities as well as the overall waiting time distribution do not follow the expected power law-like relationships.

We construct several toy models based on simple physical estimates of the relevant velocities and time scales for the waiting time distribution, and find that the observed waiting time distribution is 
consistent with both repeated failure in the crust, with the information and energy about that failure propagated solely through the crust, as well as magnetospheric reconnection, where
failure points are distributed homogeneously throughout the magnetosphere. Similarly, we find positive correlations between the spike duration and fluence as well as rise time and fluence, 
which can be explained with either a crust rupture process or magnetospheric explosive reconnection, but are inconsistent with estimates of the relationship between 
rise time and fluence in a model where crust rupture drives reconnection. 

While the theoretical estimates we make are very simple, they illustrate an important point: variability is a key property of magnetar bursts and a powerful tool for constraining the
underlying source physics. In the future, an improved interplay between theoretical models and variability studies will be crucial. We now have the tools to characterise 
variability to take advantage of 
the vast data set of magnetar bursts. As theoretical 
models evolve, they can inform the probabilistic models we build for studying variability. In return, it will be possible to test the predictions these models make using more sophisticated hierarchical models together with large samples of bursts,  and work towards an understanding of the origin and mechanisms of magnetar burst emission.

\paragraph{acknowledgements}
DH, BJB, DWH, MF and IM thank the organizers of MaxEnt2013, and also thank Ewan Cameron for valuable input to this project. DH was partially supported by a Netherlands Organization for Scientific Research (NWO) Vidi Fellowship (PI A. Watts) and the Moore-Sloan Data Science Environment at NYU. BJB is supported
by a Marsden Fast Start grant from the Royal Society of New Zealand. CE acknowledges funding from NOVA Network 4. YL acknowledges support from a Monash Reaseach Acceleration grant.  AJvdH acknowledges support from the European Research Council via Advanced Investigator Grant no. 247295 (PI: R.A.M.J. Wijers).
C.K. and was partially supported by NASA grant NNH07ZDA001-GLAST.  

\bibliography{td}

\begin{thebibliography}{}
\expandafter\ifx\csname natexlab\endcsname\relax\def\natexlab#1{#1}\fi

\bibitem[{{Andersson} {et~al.}(2009){Andersson}, {Glampedakis}, \&
  {Samuelsson}}]{andersson2009}
{Andersson}, N., {Glampedakis}, K., \& {Samuelsson}, L. 2009, \mnras, 396, 894

\bibitem[{Aschwanden(2011)}]{aschwanden2011}
Aschwanden, M. 2011, Self-Organized Criticality in Astrophysics: The Statistics
  of Nonlinear Processes in the Universe, Springer Praxis Books (Springer)

\bibitem[{{Aschwanden}(2012{\natexlab{a}})}]{aschwanden2012a}
{Aschwanden}, M.~J. 2012{\natexlab{a}}, \aap, 539, A2

\bibitem[{{Aschwanden}(2012{\natexlab{b}})}]{aschwanden2012b}
---. 2012{\natexlab{b}}, \apj, 757, 94

\bibitem[{{Aschwanden}(2013)}]{aschwanden2013b}
---. 2013, {Self-Organized Criticality Systems} (Open Academic Press)

\bibitem[{{Aschwanden} \& {Aschwanden}(2008)}]{aschwanden2008}
{Aschwanden}, M.~J., \& {Aschwanden}, P.~D. 2008, \apj, 674, 544

\bibitem[{{Aschwanden} \& {Shimizu}(2013)}]{aschwanden2013a}
{Aschwanden}, M.~J., \& {Shimizu}, T. 2013, \apj, 776, 132

\bibitem[{{Aschwanden} {et~al.}(2014){Aschwanden}, {Crosby}, {Dimitropoulou},
  {Georgoulis}, {Hergarten}, {McAteer}, {Milovanov}, {Mineshige}, {Morales},
  {Nishizuka}, {Pruessner}, {Sanchez}, {Sharma}, {Strugarek}, \&
  {Uritsky}}]{aschwanden2014}
{Aschwanden}, M.~J., {Crosby}, N., {Dimitropoulou}, M., {et~al.} 2014, ArXiv
  e-prints, arXiv:1403.6528

\bibitem[{Bak \& Chen(1989)}]{bak1989}
Bak, P., \& Chen, K. 1989, Physica D: Nonlinear Phenomena, 38, 5

\bibitem[{{Bak} {et~al.}(1987){Bak}, {Tang}, \& {Wiesenfeld}}]{bak1987}
{Bak}, P., {Tang}, C., \& {Wiesenfeld}, K. 1987, Physical Review Letters, 59,
  381

\bibitem[{{Bak} {et~al.}(1988){Bak}, {Tang}, \& {Wiesenfeld}}]{bak1988}
---. 1988, \pra, 38, 364

\bibitem[{{Baldeschi} \& {Guidorzi}(2015)}]{baldeschi2015}
{Baldeschi}, A., \& {Guidorzi}, C. 2015, \aap, 573, L7

\bibitem[{{Beloborodov} \& {Levin}(2014)}]{beloborodov2014}
{Beloborodov}, A.~M., \& {Levin}, Y. 2014, \apjl, 794, L24

\bibitem[{Bhat {et~al.}(2014)Bhat, Fishman, Briggs, Connaughton, Meegan,
  Paciesas, Wilson-Hodge, \& Xiong}]{bhat2014}
Bhat, P., Fishman, G., Briggs, M., {et~al.} 2014, Experimental Astronomy, 38,
  331

\bibitem[{Brewer {et~al.}(2011)Brewer, P{\'a}rtay, \& Cs{\'a}nyi}]{brewer2011}
Brewer, B., P{\'a}rtay, L., \& Cs{\'a}nyi, G. 2011, Statistics and Computing,
  21, 649

\bibitem[{{Brewer}(2014)}]{brewer2014}
{Brewer}, B.~J. 2014, ArXiv e-prints, arXiv:1411.3921

\bibitem[{{Brewer} {et~al.}(2013){Brewer}, {Foreman-Mackey}, \&
  {Hogg}}]{brewer2013}
{Brewer}, B.~J., {Foreman-Mackey}, D., \& {Hogg}, D.~W. 2013, \aj, 146, 7

\bibitem[{{Briggs} {et~al.}(2010){Briggs}, {Fishman}, {Connaughton}, {Bhat},
  {Paciesas}, {Preece}, {Wilson-Hodge}, {Chaplin}, {Kippen}, {von Kienlin},
  {Meegan}, {Bissaldi}, {Dwyer}, {Smith}, {Holzworth}, {Grove}, \&
  {Chekhtman}}]{briggs2010}
{Briggs}, M.~S., {Fishman}, G.~J., {Connaughton}, V., {et~al.} 2010, Journal of
  Geophysical Research (Space Physics), 115, 7323

\bibitem[{{Chaplin} {et~al.}(2013){Chaplin}, {Bhat}, {Briggs}, \&
  {Connaughton}}]{chaplin2013}
{Chaplin}, V., {Bhat}, N., {Briggs}, M.~S., \& {Connaughton}, V. 2013, Nuclear
  Instruments and Methods in Physics Research A, 717, 21

\bibitem[{{Cheng} {et~al.}(1996){Cheng}, {Epstein}, {Guyer}, \&
  {Young}}]{cheng1996}
{Cheng}, B., {Epstein}, R.~I., {Guyer}, R.~A., \& {Young}, A.~C. 1996, \nat,
  382, 518

\bibitem[{{Dib} {et~al.}(2012){Dib}, {Kaspi}, {Scholz}, \& {Gavriil}}]{dib2012}
{Dib}, R., {Kaspi}, V.~M., {Scholz}, P., \& {Gavriil}, F.~P. 2012, \apj, 748, 3

\bibitem[{{Douchin} \& {Haensel}(2001)}]{douchin2001}
{Douchin}, F., \& {Haensel}, P. 2001, \aap, 380, 151

\bibitem[{{Duncan} \& {Thompson}(1992)}]{duncan1992}
{Duncan}, R.~C., \& {Thompson}, C. 1992, \apjl, 392, L9

\bibitem[{{Esposito} {et~al.}(2010){Esposito}, {Israel}, {Turolla}, {Tiengo},
  {G{\"o}tz}, {de Luca}, {Mignani}, {Zane}, {Rea}, {Testa}, {Caraveo}, {Chaty},
  {Mattana}, {Mereghetti}, {Pellizzoni}, \& {Romano}}]{esposito2010}
{Esposito}, P., {Israel}, G.~L., {Turolla}, R., {et~al.} 2010, \mnras, 405,
  1787

\bibitem[{{Gelfand} \& {Gaensler}(2007)}]{gelfand2007}
{Gelfand}, J.~D., \& {Gaensler}, B.~M. 2007, \apj, 667, 1111

\bibitem[{{Goldreich} \& {Reisenegger}(1992)}]{goldreich1992}
{Goldreich}, P., \& {Reisenegger}, A. 1992, \apj, 395, 250

\bibitem[{{G{\"o}tz} {et~al.}(2006){G{\"o}tz}, {Mereghetti}, {Molkov},
  {Hurley}, {Mirabel}, {Sunyaev}, {Weidenspointner}, {Brandt}, {del Santo},
  {Feroci}, {G{\"o}{\u g}{\"u}{\c s}}, {von Kienlin}, {van der Klis},
  {Kouveliotou}, {Lund}, {Pizzichini}, {Ubertini}, {Winkler}, \&
  {Woods}}]{goetz2006b}
{G{\"o}tz}, D., {Mereghetti}, S., {Molkov}, S., {et~al.} 2006, \aap, 445, 313

\bibitem[{{G{\"o}{\v g}{\"u}{\c s} } {et~al.}(1999){G{\"o}{\v g}{\"u}{\c s} },
  {Woods}, {Kouveliotou}, {van Paradijs}, {Briggs}, {Duncan}, \&
  {Thompson}}]{gogus1999}
{G{\"o}{\v g}{\"u}{\c s} }, E., {Woods}, P.~M., {Kouveliotou}, C., {et~al.}
  1999, \apjl, 526, L93

\bibitem[{{G{\"o}{\v g}{\"u}{\c s}} {et~al.}(2000){G{\"o}{\v g}{\"u}{\c s}},
  {Woods}, {Kouveliotou}, {van Paradijs}, {Briggs}, {Duncan}, \&
  {Thompson}}]{gogus2000}
{G{\"o}{\v g}{\"u}{\c s}}, E., {Woods}, P.~M., {Kouveliotou}, C., {et~al.}
  2000, \apjl, 532, L121

\bibitem[{{Guidorzi}(2015)}]{guidorzi2015}
{Guidorzi}, C. 2015, Astronomy and Computing, 10, 54

\bibitem[{{Hakkila} \& {Preece}(2011)}]{hakkila2011}
{Hakkila}, J., \& {Preece}, R.~D. 2011, \apj, 740, 104

\bibitem[{{Heyl} \& {Hernquist}(2005)}]{heyl2005b}
{Heyl}, J.~S., \& {Hernquist}, L. 2005, \apj, 618, 463

\bibitem[{{Israel} {et~al.}(2008){Israel}, {Romano}, {Mangano}, {Dall'Osso},
  {Chincarini}, {Stella}, {Campana}, {Belloni}, {Tagliaferri}, {Blustin},
  {Sakamoto}, {Hurley}, {Zane}, {Moretti}, {Palmer}, {Guidorzi}, {Burrows},
  {Gehrels}, \& {Krimm}}]{israel2008}
{Israel}, G.~L., {Romano}, P., {Mangano}, V., {et~al.} 2008, \apj, 685, 1114

\bibitem[{{Israel} {et~al.}(2010){Israel}, {Esposito}, {Rea}, {Dall'Osso},
  {Senziani}, {Romano}, {Mangano}, {G{\"o}tz}, {Zane}, {Tiengo}, {Palmer},
  {Krimm}, {Gehrels}, {Mereghetti}, {Stella}, {Turolla}, {Campana}, {Perna},
  {Angelini}, \& {de Luca}}]{israel2010}
{Israel}, G.~L., {Esposito}, P., {Rea}, N., {et~al.} 2010, \mnras, 408, 1387

\bibitem[{{Jones}(2003)}]{jones2003}
{Jones}, P.~B. 2003, \apj, 595, 342

\bibitem[{{Kaneko} {et~al.}(2010){Kaneko}, {G{\"o}{\u g}{\"u}{\c s}},
  {Kouveliotou}, {Granot}, {Ramirez-Ruiz}, {van der Horst}, {Watts}, {Finger},
  {Gehrels}, {Pe'er}, {van der Klis}, {von Kienlin}, {Wachter}, {Wilson-Hodge},
  \& {Woods}}]{kaneko2010}
{Kaneko}, Y., {G{\"o}{\u g}{\"u}{\c s}}, E., {Kouveliotou}, C., {et~al.} 2010,
  \apj, 710, 1335

\bibitem[{{Kocevski} {et~al.}(2003){Kocevski}, {Ryde}, \&
  {Liang}}]{kocevski2003}
{Kocevski}, D., {Ryde}, F., \& {Liang}, E. 2003, \apj, 596, 389

\bibitem[{{Kouveliotou} {et~al.}(1998){Kouveliotou}, {Dieters}, {Strohmayer},
  {van Paradijs}, {Fishman}, {Meegan}, {Hurley}, {Kommers}, {Smith}, {Frail},
  \& {Murakami}}]{kouveliotou1998}
{Kouveliotou}, C., {Dieters}, S., {Strohmayer}, T., {et~al.} 1998, \nat, 393,
  235

\bibitem[{{Lamb} \& {Markert}(1981)}]{lamb1981}
{Lamb}, R.~C., \& {Markert}, T.~H. 1981, \apj, 244, 94

\bibitem[{{Levin} \& {Lyutikov}(2012)}]{levin2012}
{Levin}, Y., \& {Lyutikov}, M. 2012, ArXiv e-prints, arXiv:1204.2605

\bibitem[{{Li} \& {Fenimore}(1996)}]{li1996}
{Li}, H., \& {Fenimore}, E.~E. 1996, \apjl, 469, L115

\bibitem[{{Lyutikov}(2003)}]{lyutikov2003}
{Lyutikov}, M. 2003, \mnras, 346, 540

\bibitem[{{Mazets} {et~al.}(1999){Mazets}, {Aptekar}, {Butterworth}, {Cline},
  {Frederiks}, {Golenetskii}, {Hurley}, \& {Il'Inskii}}]{mazets1999}
{Mazets}, E.~P., {Aptekar}, R.~L., {Butterworth}, P.~S., {et~al.} 1999, \apjl,
  519, L151

\bibitem[{{Meegan} {et~al.}(2009){Meegan}, {Lichti}, {Bhat}, {Bissaldi},
  {Briggs}, {Connaughton}, {Diehl}, {Fishman}, {Greiner}, {Hoover}, {van der
  Horst}, {von Kienlin}, {Kippen}, {Kouveliotou}, {McBreen}, {Paciesas},
  {Preece}, {Steinle}, {Wallace}, {Wilson}, \& {Wilson-Hodge}}]{meegan2009}
{Meegan}, C., {Lichti}, G., {Bhat}, P.~N., {et~al.} 2009, \apj, 702, 791

\bibitem[{{Mereghetti}(2011)}]{mereghetti2011}
{Mereghetti}, S. 2011, Advances in Space Research, 47, 1317

\bibitem[{{Mereghetti} {et~al.}(2009){Mereghetti}, {G{\"o}tz},
  {Weidenspointner}, {von Kienlin}, {Esposito}, {Tiengo}, {Vianello}, {Israel},
  {Stella}, {Turolla}, {Rea}, \& {Zane}}]{mereghetti2009}
{Mereghetti}, S., {G{\"o}tz}, D., {Weidenspointner}, G., {et~al.} 2009, \apjl,
  696, L74

\bibitem[{{Norris} {et~al.}(1999){Norris}, {Bonnell}, \&
  {Watanabe}}]{norris1999}
{Norris}, J.~P., {Bonnell}, J.~T., \& {Watanabe}, K. 1999, \apj, 518, 901

\bibitem[{{Norris} {et~al.}(1996){Norris}, {Nemiroff}, {Bonnell}, {Scargle},
  {Kouveliotou}, {Paciesas}, {Meegan}, \& {Fishman}}]{norris1996}
{Norris}, J.~P., {Nemiroff}, R.~J., {Bonnell}, J.~T., {et~al.} 1996, \apj, 459,
  393

\bibitem[{{Piro}(2005)}]{piro2005}
{Piro}, A.~L. 2005, \apjl, 634, L153

\bibitem[{{Prieskorn} \& {Kaaret}(2012)}]{prieskorn2012}
{Prieskorn}, Z., \& {Kaaret}, P. 2012, \apj, 755, 1

\bibitem[{{Quilligan} {et~al.}(2002){Quilligan}, {McBreen}, {Hanlon},
  {McBreen}, {Hurley}, \& {Watson}}]{quilligan2002}
{Quilligan}, F., {McBreen}, B., {Hanlon}, L., {et~al.} 2002, \aap, 385, 377

\bibitem[{{Rea} {et~al.}(2014){Rea}, {Vigan{\`o}}, {Israel}, {Pons}, \&
  {Torres}}]{rea2014}
{Rea}, N., {Vigan{\`o}}, D., {Israel}, G.~L., {Pons}, J.~A., \& {Torres}, D.~F.
  2014, \apjl, 781, L17

\bibitem[{{Rea} {et~al.}(2010){Rea}, {Esposito}, {Turolla}, {Israel}, {Zane},
  {Stella}, {Mereghetti}, {Tiengo}, {G{\"o}tz}, {G{\"o}{\u g}{\"u}{\c s}}, \&
  {Kouveliotou}}]{rea2010}
{Rea}, N., {Esposito}, P., {Turolla}, R., {et~al.} 2010, Science, 330, 944

\bibitem[{{Rea} {et~al.}(2012){Rea}, {Israel}, {Esposito}, {Pons},
  {Camero-Arranz}, {Mignani}, {Turolla}, {Zane}, {Burgay}, {Possenti},
  {Campana}, {Enoto}, {Gehrels}, {G{\"o}{\v g}{\"u}{\c s}}, {G{\"o}tz},
  {Kouveliotou}, {Makishima}, {Mereghetti}, {Oates}, {Palmer}, {Perna},
  {Stella}, \& {Tiengo}}]{rea2012}
{Rea}, N., {Israel}, G.~L., {Esposito}, P., {et~al.} 2012, \apj, 754, 27

\bibitem[{{Savchenko} {et~al.}(2010){Savchenko}, {Neronov}, {Beckmann},
  {Produit}, \& {Walter}}]{savchenko2010}
{Savchenko}, V., {Neronov}, A., {Beckmann}, V., {Produit}, N., \& {Walter}, R.
  2010, \aap, 510, A77

\bibitem[{{Scargle} {et~al.}(1998){Scargle}, {Norris}, \&
  {Bonnell}}]{scargle1998}
{Scargle}, J.~D., {Norris}, J., \& {Bonnell}, J. 1998, in American Institute of
  Physics Conference Series, Vol. 428, Gamma-Ray Bursts, 4th Hunstville
  Symposium, ed. C.~A. {Meegan}, R.~D. {Preece}, \& T.~M. {Koshut}, 181--185

\bibitem[{{Scargle} {et~al.}(2013){Scargle}, {Norris}, {Jackson}, \&
  {Chiang}}]{scargle2013}
{Scargle}, J.~D., {Norris}, J.~P., {Jackson}, B., \& {Chiang}, J. 2013, \apj,
  764, 167

\bibitem[{{Scholz} \& {Kaspi}(2011)}]{scholz2011}
{Scholz}, P., \& {Kaspi}, V.~M. 2011, \apj, 739, 94

\bibitem[{{Scholz} {et~al.}(2012){Scholz}, {Ng}, {Livingstone}, {Kaspi},
  {Cumming}, \& {Archibald}}]{scholz2012}
{Scholz}, P., {Ng}, C.-Y., {Livingstone}, M.~A., {et~al.} 2012, \apj, 761, 66

\bibitem[{Skilling(1998)}]{skilling1998}
Skilling, J. 1998, in Fundamental Theories of Physics, Vol.~98, Maximum Entropy
  and Bayesian Methods, ed. G.~Erickson, J.~Rychert, \& C.~Smith (Springer
  Netherlands), 1--14

\bibitem[{Skilling(2006)}]{skilling2006}
---. 2006, Bayesian Analysis, 1, 833

\bibitem[{{Steiner} \& {Watts}(2009)}]{steiner2009}
{Steiner}, A.~W., \& {Watts}, A.~L. 2009, Physical Review Letters, 103, 181101

\bibitem[{{Strohmayer} {et~al.}(1991){Strohmayer}, {van Horn}, {Ogata},
  {Iyetomi}, \& {Ichimaru}}]{strohmayer1991}
{Strohmayer}, T., {van Horn}, H.~M., {Ogata}, S., {Iyetomi}, H., \& {Ichimaru},
  S. 1991, \apj, 375, 679

\bibitem[{{Tang} {et~al.}(2010){Tang}, {Grindlay}, {Los}, \&
  {Laycock}}]{dasch_giants}
{Tang}, S., {Grindlay}, J., {Los}, E., \& {Laycock}, S. 2010, {Astrophysical
  Journal Letters}, 710, L77

\bibitem[{{Thompson} \& {Duncan}(1995)}]{thompson1995}
{Thompson}, C., \& {Duncan}, R.~C. 1995, \mnras, 275, 255

\bibitem[{{Thompson} \& {Duncan}(2001)}]{thompson2001}
---. 2001, \apj, 561, 980

\bibitem[{{Thompson} {et~al.}(2002){Thompson}, {Lyutikov}, \&
  {Kulkarni}}]{thompson2002}
{Thompson}, C., {Lyutikov}, M., \& {Kulkarni}, S.~R. 2002, \apj, 574, 332

\bibitem[{{van der Horst} {et~al.}(2010){van der Horst}, {Connaughton},
  {Kouveliotou}, {G{\"o}{\v g}{\"u}{\c s}}, {Kaneko}, {Wachter}, {Briggs},
  {Granot}, {Ramirez-Ruiz}, {Woods}, {Aptekar}, {Barthelmy}, {Cummings},
  {Finger}, {Frederiks}, {Gehrels}, {Gelino}, {Gelino}, {Golenetskii},
  {Hurley}, {Krimm}, {Mazets}, {McEnery}, {Meegan}, {Oleynik}, {Palmer},
  {Pal'shin}, {Pe'er}, {Svinkin}, {Ulanov}, {van der Klis}, {von Kienlin},
  {Watts}, \& {Wilson-Hodge}}]{vanderhorst2010}
{van der Horst}, A.~J., {Connaughton}, V., {Kouveliotou}, C., {et~al.} 2010,
  \apjl, 711, L1

\bibitem[{{van der Horst} {et~al.}(2012){van der Horst}, {Kouveliotou},
  {Gorgone}, {Kaneko}, {Baring}, {Guiriec}, {G{\"o}{\v g}{\"u}{\c s}},
  {Granot}, {Watts}, {Lin}, {Bhat}, {Bissaldi}, {Chaplin}, {Finger}, {Gehrels},
  {Gibby}, {Giles}, {Goldstein}, {Gruber}, {Harding}, {Kaper}, {von Kienlin},
  {van der Klis}, {McBreen}, {Mcenery}, {Meegan}, {Paciesas}, {Pe'er},
  {Preece}, {Ramirez-Ruiz}, {Rau}, {Wachter}, {Wilson-Hodge}, {Woods}, \&
  {Wijers}}]{vanderhorst2012}
{van der Horst}, A.~J., {Kouveliotou}, C., {Gorgone}, N.~M., {et~al.} 2012,
  \apj, 749, 122

\bibitem[{{van der Klis}(2006)}]{xrb_khzqpos}
{van der Klis}, M. 2006, Advances in Space Research, 38, 2675

\bibitem[{{van Hoven} \& {Levin}(2008)}]{vanhoven2008}
{van Hoven}, M., \& {Levin}, Y. 2008, \mnras, 391, 283

\bibitem[{{von Kienlin} {et~al.}(2012){von Kienlin}, {Gruber}, {Kouveliotou},
  {Granot}, {Baring}, {G{\"o}{\v g}{\"u}{\c s}}, {Huppenkothen}, {Kaneko},
  {Lin}, {Watts}, {Bhat}, {Guiriec}, {van der Horst}, {Bissaldi}, {Greiner},
  {Meegan}, {Paciesas}, {Preece}, \& {Rau}}]{vonkienlin2012}
{von Kienlin}, A., {Gruber}, D., {Kouveliotou}, C., {et~al.} 2012, \apj, 755,
  150

\bibitem[{{Woods} \& {Thompson}(2006)}]{woods2006}
{Woods}, P.~M., \& {Thompson}, C. 2006, {Soft gamma repeaters and anomalous
  X-ray pulsars: magnetar candidates}, ed. W.~H.~G. {Lewin} \& M.~{van der
  Klis}, 547--586

\bibitem[{{Younes} {et~al.}(2014){Younes}, {Kouveliotou}, {van der Horst},
  {Baring}, {Granot}, {Watts}, {Bhat}, {Collazzi}, {Gehrels}, {Gorgone},
  {G{\"o}{\u g}{\"u}{\c s}}, {Gruber}, {Grunblatt}, {Huppenkothen}, {Kaneko},
  {von Kienlin}, {van der Klis}, {Lin}, {Mcenery}, {van Putten}, \&
  {Wijers}}]{younes2013}
{Younes}, G., {Kouveliotou}, C., {van der Horst}, A.~J., {et~al.} 2014, \apj,
  785, 52

\end{thebibliography}
\bibliographystyle{apj}

\end{document}